\documentclass[iop,revtex4]{emulateapj}

\usepackage{amsmath}
\usepackage{amssymb}
\usepackage{graphicx}
\usepackage{multirow}
\usepackage{enumerate}
\usepackage{threeparttable}
\usepackage[breaklinks,colorlinks,citecolor=blue]{hyperref}

\slugcomment{160221 version}

\begin{document}

\title{Magnetic fields in the massive dense cores of DR21 filament: weakly magnetized cores in a strongly magnetized filament}
\author{Tao-Chung Ching\altaffilmark{1,2}, Shih-Ping Lai\altaffilmark{1,3}, Qizhou Zhang\altaffilmark{2}, Josep M.\ Girart\altaffilmark{4,2}, Keping Qiu\altaffilmark{5,6}, and Hauyu B.\ Liu\altaffilmark{7}}
\email{chingtaochung@gmail.com}

\altaffiltext{1}{Institute of Astronomy and Department of Physics, National Tsing Hua University, Hsinchu 30013, Taiwan}
\altaffiltext{2}{Harvard-Smithsonian Center for Astrophysics, 60 Garden Street, Cambridge MA 02138, USA}
\altaffiltext{3}{Institute of Astronomy and Astrophysics, Academia Sinica, P.O. Box 23-141, Taipei 10617, Taiwan}
\altaffiltext{4}{Institut de Ci\`{e}ncies de l'Espai, (CSIC-IEEC), Campus UAB, Carrer de Can Magrans, S/N, 08193 Cerdanyola del Vall\`es, Catalonia, Spain}
\altaffiltext{5}{School of Astronomy and Space Science, Nanjing University, 163 Xianlin Avenue, Nanjing 210023, China}
\altaffiltext{6}{Key Laboratory of Modern Astronomy and Astrophysics (Nanjing University), Ministry of Education, Nanjing 210023, China}
\altaffiltext{7}{European Southern Observatory (ESO), Karl-Schwarzschild-Str. 2, D-85748 Garching, Germany}

\begin{abstract}
We present Submillimeter Array 880 $\mu$m dust polarization observations of six massive dense cores in the DR21 filament. 
The dust polarization shows complex magnetic field structures in the massive dense cores with sizes of 0.1 pc, in contrast to the ordered magnetic fields of the parsec-scale filament. 
The major axes of the massive dense cores appear to be aligned either parallel or perpendicular to the magnetic fields of the filament, indicating that the parsec-scale magnetic fields play an important role in the formation of the massive dense cores. However, the correlation between the major axes of the cores and the magnetic fields of the cores is less significant, suggesting that during the core formation, the magnetic fields below 0.1 pc scales become less important than the magnetic fields above 0.1 pc scales in supporting a core against gravity.
Our analysis of the angular dispersion functions of the observed polarization segments yields the plane-of-sky magnetic field strengths of 0.4--1.7 mG of the massive dense cores. We estimate the kinematic, magnetic, and gravitational virial parameters of the filament and the cores.
The virial parameters show that in the filament, the gravitational energy is dominant over magnetic and kinematic energies, while in the cores, the kinematic energy is dominant. Our work suggests that although magnetic fields may play an important role in a collapsing filament, the kinematics arising from gravitational collapse must become more important than magnetic fields during the evolution from filaments to massive dense cores.
\end{abstract}


\section{Introduction}
Resent observations of molecular clouds have revealed a two-step scenario of star formation that supersonic flows first compress clouds into parsec-long filaments, and then gravitational bound filaments collapse and fragment into dense cores\footnote{This paper follows the nomenclature used in \citet{2009Zhang}, and refers to a cloud as an entity of molecular gas of 10--100 pc, a clump or filament as an entity of few pc that forms massive stars with a population of lower mass stars, and dense cores as an entity of 0.01--0.1 pc that forms one or a group of stars.} \citep{2014Andre}.
In this scenario, massive dense cores ($\sim$ 0.1 pc, $\ga$ 20 M$_{\sun}$) formed at intersections of filaments are the sites of clustered and massive star formation 
\citep{2010GM, 2012Hennemann, 2012Schneider, 2012Liu, 2013Peretto}.
Massive dense cores 
are magnetized with magnetic field strengths of 0.1--1 mG \citep{2008Falgarone, 2013Girart, 2013Qiu, 2014Frau, 2015Li, 2016Houde}.
Since the star-formation rate in dense gas is only few percents of the rate of a free-fall collapse, magnetic fields are suggested to be an important ingredient supporting massive dense cores against self-gravity \citep{2007KT}.

The orientation between cloud morphology and magnetic fields reveal information about the role of magnetic fields in star formation.
In a strongly magnetized cloud, ions are easier to move along the field lines than perpendicular to the field lines.
Simulations of magnetized turbulent gas show that if the turbulent energy is stronger than the gravitational energy, turbulence extends gas along the field lines and forms filaments parallel to magnetic fields \citep[see Figure 2 of][]{1998Stone}. 
On the other hand, if the gravitational energy is stronger than the turbulent energy, the mass contraction along the field lines results in filaments perpendicular to magnetic fields \citep{2008NL,2016Li}.
That is, strong magnetic field models predict either-parallel-or-perpendicular alignment between filaments and magnetic fields, depending on the competition between turbulent pressure and gravitational pressure.
In contrast, the filaments formed with weak magnetic fields tends to be randomly aligned with respect to magnetic fields \citep{1998Stone,2016Li}.
Such alignment of strong magnetic field models and misalignment of weak magnetic field models are also presented in the simulations not including turbulence \citep{2014VanLoo}.
In observations, the correlation between the orientations of clouds and magnetic fields is found to be either-parallel-or-perpendicular in filamentary clouds from few to tens pc scales \citep{2013Li,2016Planck} and in massive cores at 0.1--0.01 pc scales \citep{2014Zhang,2014Koch}, suggesting a dynamically important role of magnetic fields in various physical scales (see \citeauthor{2014Li} \citeyear{2014Li} for a review).

Observations of the polarized dust emission from magnetically aligned dust grains have been proven to be the most efficient way to reveal magnetic field structures in molecular clouds \citep{2012Crutcher}. For massive dense cores with typical sizes of 0.1 pc and distances of few kiloparsecs, several interferometric observations have been performed to spatially resolve the magnetic field structures \citep[][also see Zhang et al.\ 2014 for a review of SMA polarization legacy project]{1998Rao,2001Lai,2002Lai,2003Lai,2008Cortes,2016Cortes,2009Girart,2013Girart,2009aTang,2009bTang, 2010Tang, 2013Tang, 2013Liu, 2013Qiu, 2014Qiu, 2014Hull,2014Frau,2014Sridharan,2015Li}.  
The observations reveal diverse magnetic field structures from ordered hour-glass shapes (e.g.\ G31.41, \citeauthor{2009Girart} \citeyear{2009Girart}) to chaotic distributions (e.g.\ G5.89,  \citeauthor{2009aTang} \citeyear{2009aTang}). 
The magnetic energy in a massive core is typically equal or less than the energies of rotation, outflows, and radiation \citep{2009aTang,2013Girart,2013Qiu,2014Frau,2014Sridharan}. 

In this paper, we present a Submillimeter Array (SMA) study of six massive dense cores in the DR21 filament.
DR21 filament is the densest and most massive region in the Cygnus X complex \citep{2007Motte,2011Roy,2016Schneider}, hosting several well-studied high-mass star-forming regions like DR21 and DR21(OH) \citep{1966DR} at a distance of 1.4 kpc \citep{2012Rygl}.
{\it Herschel} observations showed that the ridge of DR21 filament has a length of 4 pc and a mass of 15000 M$_{\sun}$, connected by several sub-filaments with masses between 130 to 1400 M$_{\sun}$ \citep{2012Hennemann}. Molecular line studies suggested continuous mass flows from the sub-filaments onto the DR 21 filament, driving the global gravitational collapse of the filament \citep{2010Schneider}. 
Dust continuum studies found 12 massive dense cores in the DR21 filament \citep{2007Motte}.
The massive dense cores drive active outflows \citep{2007Davis,2007Motte} and masers \citep{1983BE,2000Argon,2005Pestalozzi}, indicating recent high-mass star formation.
Thirty-two fragments with sizes of a few thousands AU were detected in a sample of seven cores, whereas some of the fragments were proposed to be precursors of OB stars \citep{2010Bontemps,2012Zapata}.
Interferometric molecular line observations reveal small-scale converging flows within the massive dense cores, which may initiate the dense structures of the cores \citep{2011aCsengeri,2011bCsengeri}.

Magnetic fields of the DR21 filament at parsec scale have been mapped through single-dish polarimetric observations \citep{1994MM,1999Glenn,2006VF,2009Kirby}, revealing an uniform structure of magnetic fields aligned perpendicular to the filament.
Zeeman observations of the CN lines found a 0.4--0.7 mG magnetic field strength of the line-of-sight component in DR21(OH) \citep{1999Crutcheretal,2008Falgarone}.
In contrast to the uniform magnetic fields at pc scales, interferometric dust polarization observations revealed complex magnetic fields in the DR21(OH) \citep{2003Lai,2013Girart}.  
Here, we present SMA dust polarization observations of massive dense cores Cyg-N38 (alias DR21(OH)-W), Cyg-N43 (W75S-FIR1), Cyg-N44 (DR21(OH)), Cyg-N48 (DR21(OH)-S), Cyg-N51 (W75S-FIR2), and Cyg-N53. 
The molecular line data of our observations will be represented in our upcoming paper (Paper II, Ching et al.\ in prep.) Section 2 describes the observations and data reduction. Section 3 presents the dust polarization maps. Analysis and discussion of the core property and magnetic fields are presented in Section 4. In Section 5, we draw the main conclusions.

\section{Observations and Data Reduction}\label{sec_obs}
The SMA\footnote{The Submillimeter Array is a joint project between the Smithsonian Astrophysical Observatory and the Academia Sinica Institute of Astronomy and Astrophysics and is funded by the Smithsonian Institution and the Academia Sinica.} polarization \citep{2004Ho,2006Marrone} observations were carried out toward six massive dense cores in the DR21 filament from 2011 to 2015 in the 345 GHz band. The observations of the Cyg-N44 were previously published by \citet{2013Girart} and \citet{2014Zhang}.
Table \ref{obs_table} lists the dates, array configurations, number of antennas, polarimeter mode, and calibrators of the observations of the other five sources. 
The pointing centers of the sources are listed in Table \ref{map_table}.
The sources were part of the sample of the SMA polarization legacy project and were not reported in \citet{2014Zhang} except Cyg-N44. The observations before 2013 were carried out in the single receiver mode, and the observations after 2013 were carried out with dual receivers. 
The dual-receiver mode offers a $\sqrt{2}$ improvement in the signal-to-noise ratio (S/N) of continuum polarization measurement than the single-receiver mode.
The bandpass calibrators were also served as the polarization calibrator in each observation. 
In some observations, when the data of bandpass calibrators were not good enough for polarization calibration, we adopted the leakage table from adjacent days to calibrate polarization.
The absolute flux was determined from observations of planets or planetary moons, and the typical flux uncertainty in SMA observations was estimated to be $\sim$ 20$\%$.

The data were calibrated in the IDL MIR package for flux, bandpass and time-dependent gain, and were exported to MIRIAD for further processing.
The intrinsic instrumental polarizations (i.e., leakage) of the lower sideband and the upper sideband were calibrated independently with the MIRIAD task GPCAL to an accuracy of 0.1$\%$ \citep{2008MR} and were removed from the data. 
The continuum visibilities were averaged from channels absent of line emission to produce the Stokes $I$, $Q$, and $U$ maps. 
Self-calibration for sources with strong Stokes $I$ continuum emission (S/N $\ga$ 60) was performed to refine the gain solutions. 
The Stokes $I$, $Q$, and $U$ maps were independently deconvolved using the CLEAN algorithm. 
Finally, the polarized intensity, position angle, and polarization percentage along with their uncertainties were derived from the Stokes $I$, $Q$, and $U$ maps using the MIRIAD task IMPOL.

Table \ref{map_table} lists the synthesized beams and rms noises of the maps presented in this paper. 
The maps were made with the natural weighting to maximize the sensitivity.
For the Cyg-N43 and Cyg-N51 data, we applied a Gaussian taper of 2$\arcsec$ FWHM to enhance the polarization detection.
The baselines spanned from 9 k$\lambda$ to 85 k$\lambda$, providing a resolution of 3$\arcsec$--4$\arcsec$ and a largest scale of 10$\arcsec$ with 50\% flux recovery \citep{1994WW}.
The on-source time is at least 3 hour per target, resulting in a rms noise level less then 2 mJy beam$^{-1}$ in the Stokes $Q$ and $U$ maps.
Due to the limited dynamic range of the SMA, the rms noise of the continuum Stokes $I$ emission was significantly higher than that of the Stokes $Q$ and $U$ emission. 
In this paper, we set a polarization cutoff level at S/N equal to 3.
The majority of polarization detections have S/N better than 5, thus the typical uncertainty in the polarization angle is $\sim$ 6$\arcdeg$ \citep{1993NC}. 

To illustrate the magnetic fields of the DR21 filament, we also present the polarization observations of James Clerk Maxwell Telescope (JCMT). We obtained the data from the SCUBA Polarimeter Legacy Catalog compiled by \citet{2009Matthews}, which are sampled on a 10$\arcsec$ grid with a resolution of 20$\arcsec$. The noise level of the JCMT Stokes $Q$ and $U$ data is $\sim$ 10 mJy beam$^{-1}$.

\section{Results}\label{sec_result}
\subsection{JCMT and SMA polarization maps}
\begin{figure}
\includegraphics[scale=0.8]{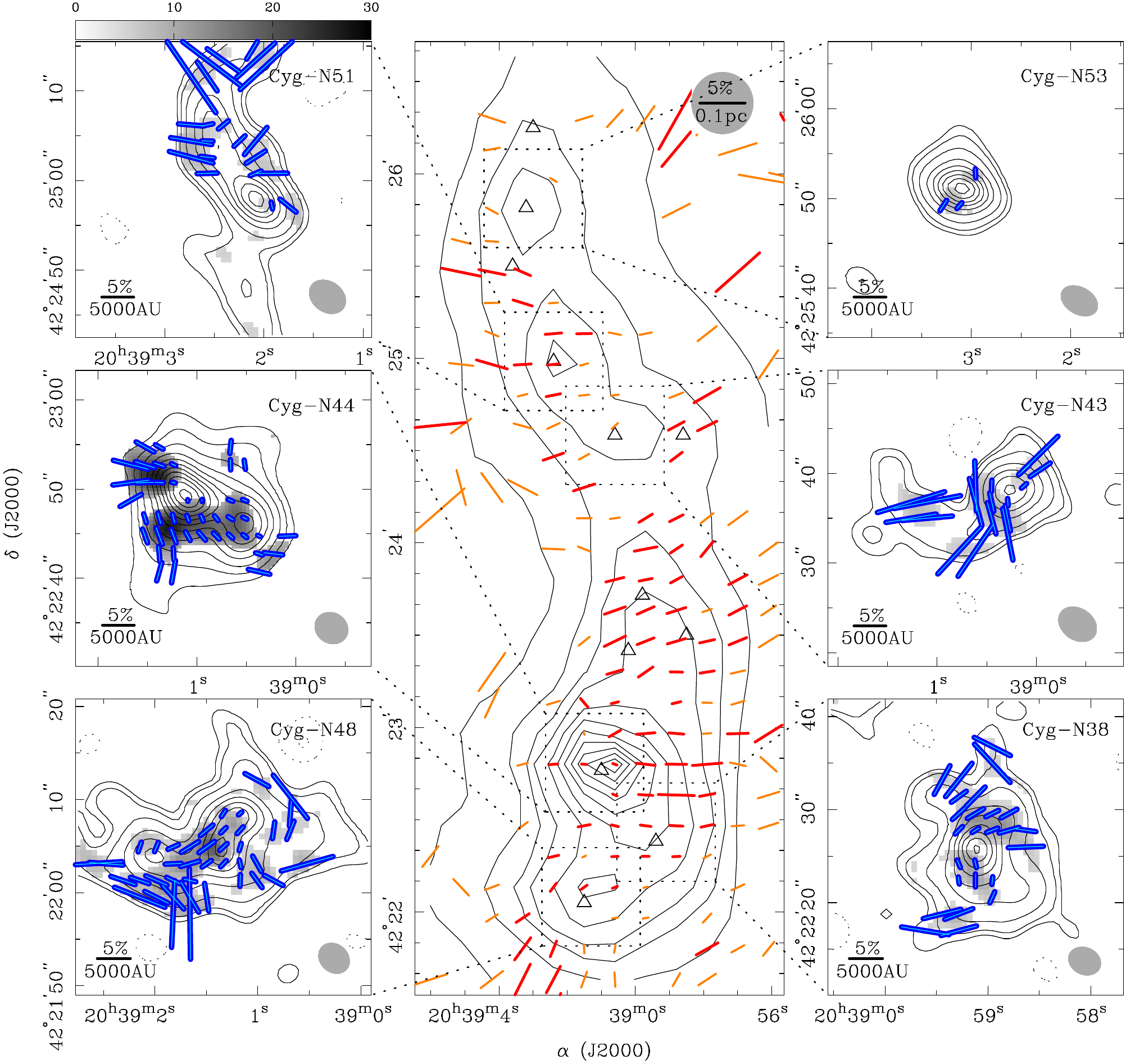}
\caption{Middle column: the JCMT SCUBA dust polarization map at 850 $\mu$m toward the DR21 filament \citep{2006VF, 2009Matthews}. The contours show the total dust continuum emission with levels of 5\%, 10\%, 15\%, 20\%, 30\%,..., 90\% of the peak. The red and orange segments represent the polarization detections at S/N above 3 and between 2--3, respectively. 
The segments show the magnetic field direction with its length proportional to the polarization percentage.
The triangles mark the positions of the millimeter sources in \citet{2007Motte}. 
Left and right panels: the SMA dust polarization maps at 880 $\mu$m toward six massive cores in the DR21 filament. In the maps of Cyg-N38, Cyg-N43, Cyg-N48, Cyg-N51, and Cyg-N53, the contour levels are -5, -3, 3, 5, 10, 20,..., 70 times the rms noises of the maps. The Cyg-N44 map is the SMA subcompact, compact and extended map in \citet{2013Girart}, which shows contours at 4\%, 8\%, 18\%, 28\%,..., 98\% of the peak. The polarized intensity is shown in gray image with the scale bar at top-left corner of the figure in units of Jy beam$^{-1}$. 
The blue segments show the SMA polarizations of S/N $\geq$ 3 with the directions of magnetic fields and the lengths proportional to the polarization percentage.
In each panel, the synthesized beam is plotted as a filled grey ellipse.
The size of the SMA panels is 32$\arcsec$, equal to the diameter of SMA primary beam at 880 $\mu$m.}
\label{fig_dustpol}
\end{figure}

Figure \ref{fig_dustpol} presents the SMA 880 $\mu$m and JCMT 850 $\mu$m polarization maps of the DR21 filament and six massive dense cores. 
The JCMT map shows that the DR21 filament is elongated in the north--south direction.  
The northern part of the filament is tiled in the northeast direction, and the most southern tip of the filament is tiled in the southeast direction.
The magnetic fields of the DR21 filament are mainly in the east--west direction, and thus almost perpendicular to the filament.
The variation in the direction of magnetic fields is relatively small, except that in the most southern tip of the filament, the magnetic fields are significantly changed to the southeast--northwest direction and become parallel to the filament. More detailed studies of the JCMT polarization map of DR21 filament are represented in \citet{2006VF} and \citet{2013Girart}.

The SMA maps at 3$\arcsec$--4$\arcsec$ ($\sim$ 0.03 pc or 5000 AU) resolution reveal compact structures of the massive dense cores. 
Owing to the filtering effect of interferometers, the SMA is more sensitive to the small-scale magnetic fields in the cores, whereas the JCMT is more sensitive to the large-scale magnetic field of the filament.
In contrast to the relatively uniform magnetic fields of the filament, the SMA maps display complex structures of the magnetic fields of the cores. 
In addition, the SMA maps show significant depolarization effect toward the center of core, which could be resulted from the distortion of magnetic fields at high column densities \citep{2016Chen}. 
The dust emission and magnetic fields of Cyg-N44 have been studied in several works \citep{1989Woody,2003Lai,2012Zapata,2013Girart,2014Zhang}. 
Here we focus on the SMA maps of the other five massive dense cores.

\noindent $\bullet$ Cyg-N38: The dust emission of Cyg-N38 shows a north--south elongated peak, consistent with the N$_{2}$H$^{+}$ 1--0 emission of the core \citep{2011bCsengeri}. In contrast to the east-west orientated JCMT polarization segment, the SMA polarization segments of Cyg-N38 show northwest-southeast oriented magnetic fields in the northern and southern parts of the core and north-south oriented magnetic fields in the center. The polarized emission is weak at the gaps between the northwest-southeast oriented and north-south oriented magnetic fields. The depolarization gaps might be caused by a perturbation in dust alignment or by an overlap of two components of magnetic fields with different orientations. Since our molecular line data show that the line-of-sight velocity in the northern and southern parts of the core is different from that in the center of the core (Paper II), the northwest-southeast oriented and the north-south oriented segments may belong to two independent components of magnetic fields.

\noindent $\bullet$ Cyg-N43: The dust emission of Cyg-N43 shows an arc-like structure with the peak emission concentrated at the western end of the arc. Similar to Cyg-N38, the polarization segments of Cyg-N43 show one group of north-south oriented magnetic fields in the center of the core, and one group of northwest-southeast oriented magnetic fields in the northern and southern parts of the core. Depolarization gaps exist at the places between the two groups of polarization segments, again suggesting that the north-south magnetic fields and the northwest-southeast magnetic fields may not be directly connected to each other.

\noindent $\bullet$ Cyg-N48: Cyg-N48 has the most extended dust emission in the DR21 filament. The SMA map shows that Cyg-N48 is composed of several sources, similar to the ``S'' shaped morphology of the 3 mm continuum map \citep{2010Bontemps}. Cyg-N48 also has the most extended polarized emission. The magnetic fields of Cyg-N48 rotate smoothly from the northwest-southeast direction in the center of the core to the east-west direction in the east of the core. The change of the magnetic fields from the northwest-southeast direction to the east-west direction appears to be consistent with the curvature of the ``S'' shaped morphology of the core.

\noindent $\bullet$ Cyg-N51: Cyg-N51 is significantly elongated along the north--south direction, following the large scale filament. The direction of magnetic fields in Cyg-N51 is similar to the east--west direction of magnetic fields in the filament, revealing that the orientation between magnetic fields and dust emission is consistently perpendicular from filament to core. Above the fifth contour in the continuum emission, very few polarization segments are detected, suggesting that the magnetic fields in the center of Cyg-N51 could be different from the east--west fields revealed by the SMA and JCMT maps.

\noindent $\bullet$ Cyg-N53: Cyg-N53 is only partially resolved and hence has a round shape with a centrally concentrated peak. Although Cyg-N53 appears to be a single object in our SMA map, 7 fragments in Cyg-N53 with sizes of few thousands AU have been found in the 1 mm continuum map with higher resolution \citep{2010Bontemps}.  
Compared to the other cores, Cyg-N53 has the weakest polarized emission. The weak polarized emission in the SMA map is consistent with the weak polarized emission in the JCMT map. 

\subsection{Combined JCMT and SMA 880 $\mu$m continuum maps}
\begin{figure}
\includegraphics[scale=0.7]{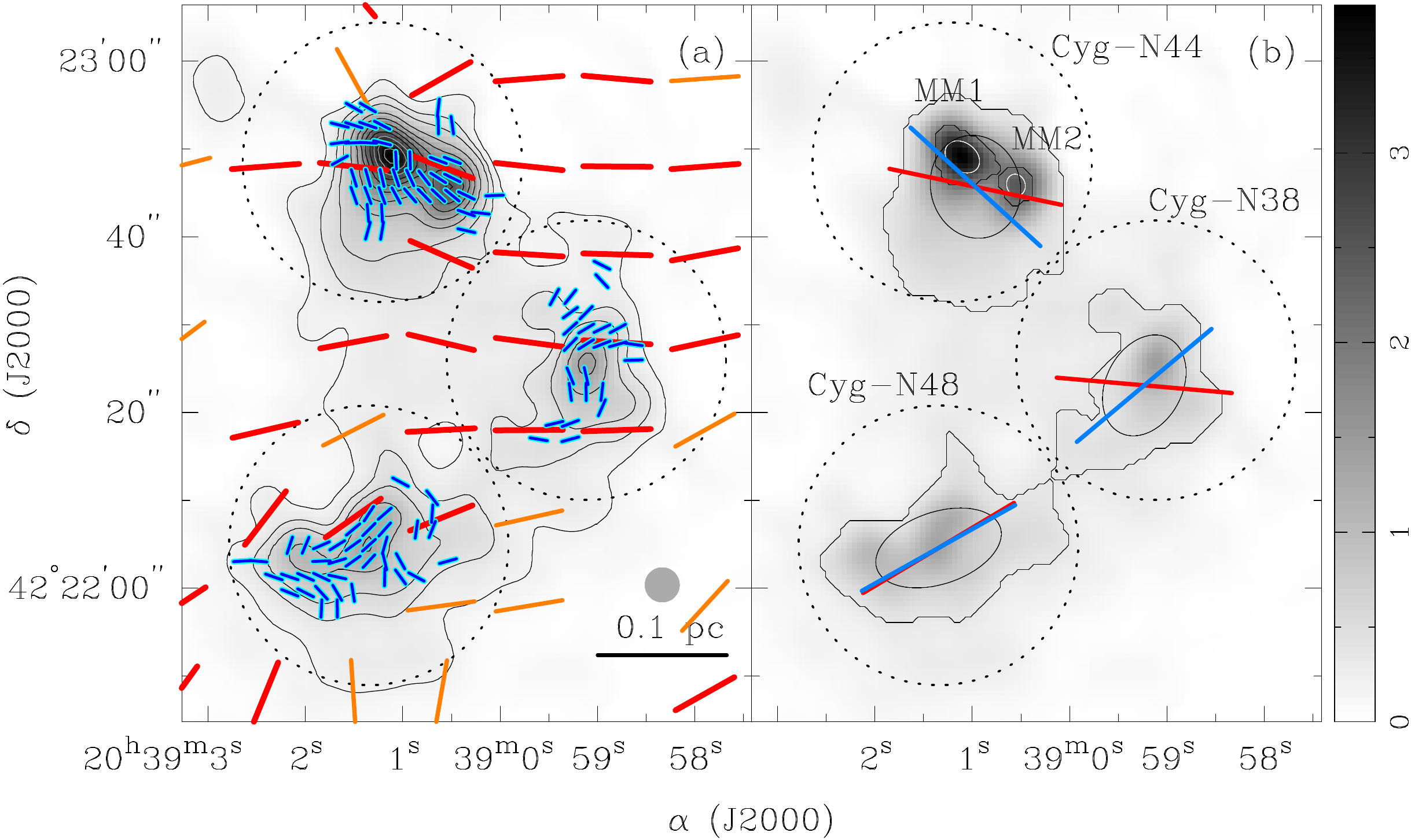}
\caption{JCMT and SMA combined 880 $\mu$m continuum maps of the Cyg-N38, Cyg-N44, and Cyg-N48.
(a) The gray scale in units of Jy beam$^{-1}$ and the contours show the combined dust Stokes $I$ emission. The contours are plotted at 5\%, 10\%, 15\%, 25\%,..., 95\% of the peak.
The dotted circles represent the SMA primary beams of the sources.
The synthesized beam of the combined map is plotted as a filled grey circle.
The red, orange, and blue segments are the segments in Figure 1 with equal length (please note that we do NOT combine the JCMT and SMA Stokes $Q$ and $U$ maps).  
(b) The structures selected by dendrograms overlaid on the JCMT and SMA combined map.
The contours show the boundries of the structures, and the ellipses are plotted using the major FWHM, minor FWHM, and $\theta_{core}$ computed by dendrograms.
For each source, the red (or orange if S/N between 2--3) and blue segments show the orientations of $\theta_{JCMT}$ and $\theta_{SMA}$, respectively.}
\label{fig_jcmtsma_south}
\end{figure}

\begin{figure}
\includegraphics[scale=0.7]{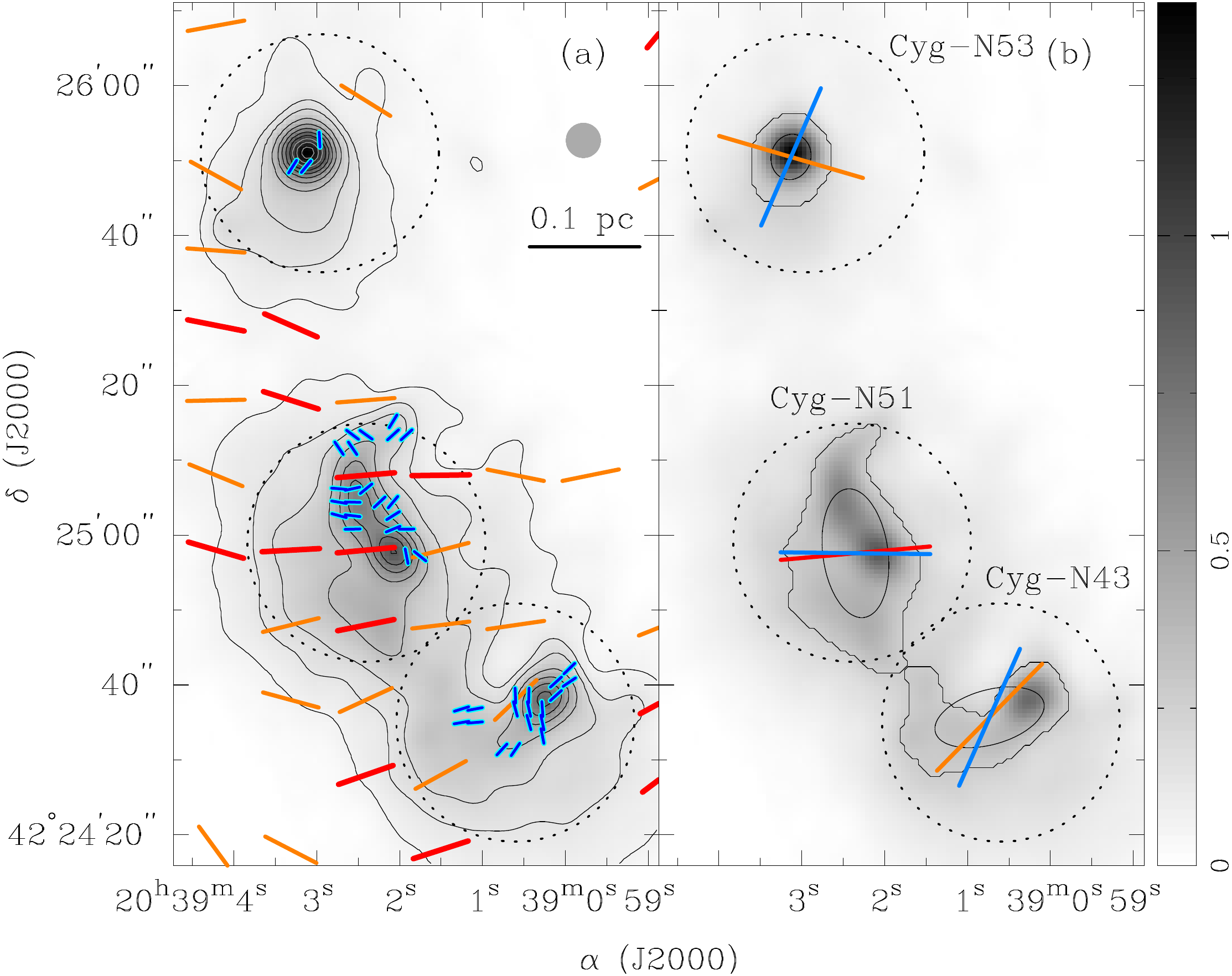}
\caption{JCMT and SMA combined 880 $\mu$m continuum maps of the Cyg-N43, Cyg-N51, and Cyg-N53. The legends are the same as in Figure 2.}
\label{fig_jcmtsma_north}
\end{figure}

Figures \ref{fig_jcmtsma_south} and \ref{fig_jcmtsma_north} present the JCMT\footnote{
When combined the JCMT and SMA Stokes $I$ maps, we used the SCUBA2  commissioning data (Proposal ID M11BEC30) from the JCMT Science Archive, which are sampled on a 4$\arcsec$ grid with a resolution of 15$\arcsec$, better than the SCUBA Pol data.}
and SMA combined continuum maps at 880 $\mu$m of the Cyg-N38, Cyg-N44, and Cyg-N48 in the south of the filament and the Cyg-N43, Cyg-N51, and Cyg-N53 in the north of the filament, respectively. To produce the combined maps, we first scaled the JCMT map at 850 $\mu$m to the wavelength of 880 $\mu$m assuming a dust emissivity index of 2.0, and then combined the scaled JCMT map and SMA maps using the FEATHER task in CASA and the LINMOS task in MIRIAD. The resulted resolution and noise level of the combined Cyg-N38, Cyg-N44, and Cyg-N48 map are 3.9$\arcsec$ and 40 mJy beam$^{-1}$. The resulted resolution and noise level of the combined Cyg-N43, Cyg-N51, and Cyg-N53 map are 4.6$\arcsec$ and 30 mJy beam$^{-1}$.
The noise levels of the combined maps are mainly contributed from the SCUBA2 data.
 
Figure \ref{fig_jcmtsma_south} shows that the northwestern edge of Cyg-N48 is connected to the southeastern edge of Cyg-N38, and the northeastern edge of Cyg-N38 is connected to the southwestern edge of Cyg-N44. 
Since the emission between Cyg-N44 and Cyg-N48 is fainter than the emission between Cyg-N38 and Cyg-N44 and the emission between Cyg-N38 and Cyg-N48, the three cores seem to be aligned in a bow-like structure. This bow-like structure connecting the three cores are also presented in the H$^{13}$CO$^+$ and N$_2$H$^+$ maps of the DR21 filament \citep{2011aCsengeri,2011bCsengeri}. 
Figure \ref{fig_jcmtsma_north} shows that Cyg-N43, Cyg-N51, and Cyg-N53 are aligned on the northern ridge of the DR21 filament. The eastern edge of Cyg-N43 is connected to the southern edge of Cyg-N51. The emission between Cyg-N51 and Cyg-N53 is weak in the JCMT map, suggesting that Cyg-N53 is more isolated than the other cores.
 
\subsection{Masses, Densities, and Sizes of the Cores}
We performed dendrograms \citep{2008Rosolowsky,2009Goodman} using the {\it astrodendro} software\footnote{http://www.dendrograms.org} on the JCMT and SMA combined maps to characterize the structures of the cores. 
Since all the cores are located on the ridge of the filament, we chose a lower limit of 7$\sigma$ for the dendrogram to detect a structure.
The required difference to separate structures of the dendrogram was set to be 1$\sigma$, and the minimum area of a structure was one synthesis beam.
Figures \ref{fig_jcmtsma_south}b and \ref{fig_jcmtsma_north}b show the structures identified by the dendrogram. The FWHM sizes along the major axes, the FWHM sizes along the minor axes, the mean sizes ($FWHM_{mean}$), and the position angles ($\theta_{core}$) of the structures computed by the dendrogram are listed in Table \ref{phy_table}, along with the integrated flux of the structures after a subtraction of the background emission below the cutoffs of the structures.
To estimate the errors in the FWHM sizes and in $\theta_{core}$, we performed bootstrap method \citep{2002Press} by adding additional observational noises to the JCMT and SMA combined maps and running dendrograms on those simulated maps. The procedure was repeated for 100 times, and the errors were estimated from the sample of 
100 FWHM sizes and 100 $\theta_{core}$ identified by dendrograms.
The size and flux of Cyg-N44 selected by the dendrogram are 11.1$\arcsec$ and 21.4 Jy, compatible with the results of 10.4$\arcsec$ and 18.8 Jy derived from the SMA map \citep{2013Girart}. 
The bright sources MM1 and MM2 in Cyg-N44 \citep{1989Woody} can also be separated by the dendrogram.
Although Cyg-N43, Cyg-N48, and Cyg-N51 show clear multiple intensity peaks in the SMA maps, the dendrogram failed to identify fragments of the sources owing to the resolution of the combined maps.

Under the assumptions of optical thin emission and isothermal core, the source mass $M$ is related to the integrated flux $S_{int}$ as
\begin{equation}
M = \frac{ \gamma S_{int} d^{2}}{\kappa B(T)}, 
\end{equation}
where $\gamma$ is the gas-to-dust ratio, $d$ is the distance of the source, $\kappa$ is the dust opacity, and $B(T)$ is the Planck function at a dust temperature $T$.
We adopt a gas-to-dust ratio of 100, a distance of 1.4 kpc, and a dust opacity of 1.5 cm$^2$ g$^{-1}$ of the cool (10--30 K) and dense ($\sim$ 10$^6$ cm$^{-3}$ ) dust mantles \citep{1994OH}.  
We adopt a temperature of 20 K for the cores since Herschel maps show a $\sim$ 16--22 K dust temperature of the cores in the DR21 filament \citep{2012Hennemann}, except a higher temperature of 30 K for the hot core Cyg-N44 \citep{1973Mayer}. 
Assuming a radius equal to the $FWHM_{mean}$ of the source \citep{2007Motte}, the averaged column and volume density of the source are estimated as follows:
\begin{equation}
N_{H_2} = \frac{M}{\pi \times FWHM_{mean}^2}, 
\end{equation}
\begin{equation}
n_{H_2} = \frac{M}{\frac{4}{3} \pi \times FWHM_{mean}^3}.
\end{equation}

The derived masses, column densities, and volume densities are listed in Table \ref{phy_table}, along with the errors derived from the propagation of the errors in $S_{int}$ and $FWHM_{mean}$.
Although the derived errors are typically smaller than 10\%, the uncertainties in the flux calibration, gas-to-dust ratio, distance, dust emissivity and more critically, the temperature may lead to an overall 30\%--50\% uncertainty.
The masses estimated from our JCMT and SMA combined maps are about a factor of two to four lower than those estimated from single-dish observations \citep{2007Motte,2008Di,2013Palau}, mainly because we performed dendrograms above a cutoff of 7$\sigma$ to separating cores and therefore the core sizes estimated in this work are about half of the sizes in single-dish maps.

 \subsection{Large-scale and Small-scale Magnetic Fields} 
\begin{figure}[!b]
\includegraphics[scale=1]{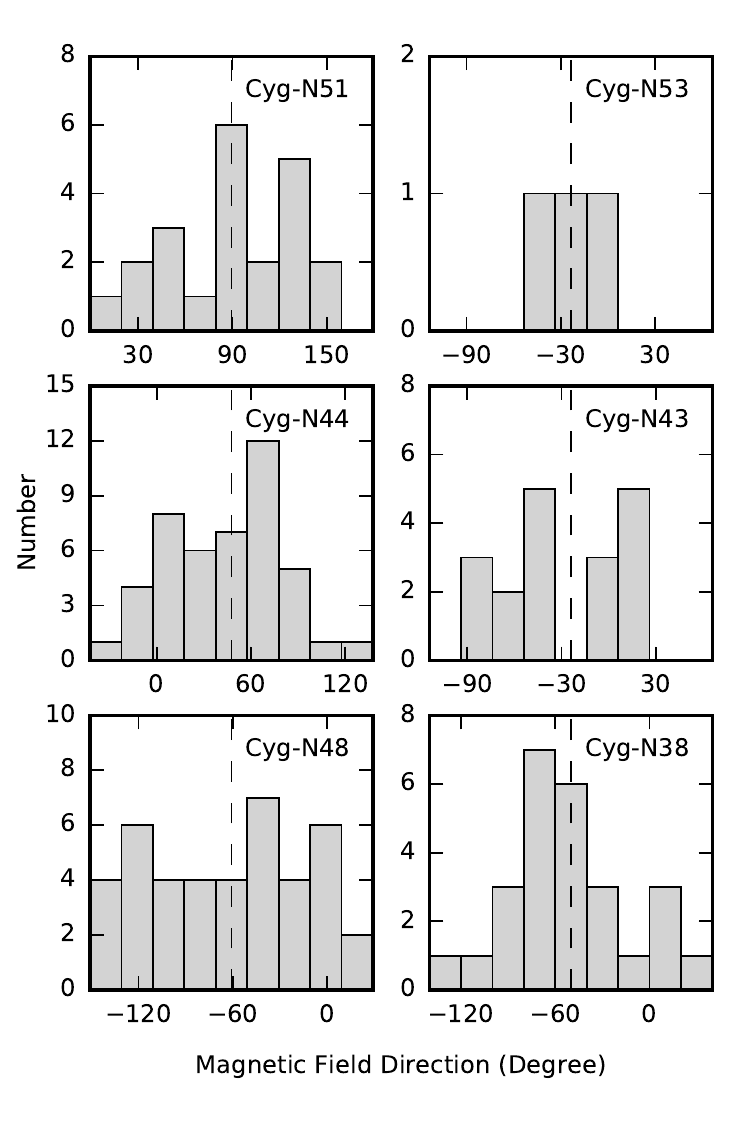}
\caption{Distributions of the magnetic field orientations inferred from the SMA polarization detections. A dashed line shows the averaged orientation of magnetic field inferred from the uncertainty-weighted average of the SMA polarization angles of each source. The order of the panels is arranged as the panels in Figure \ref{fig_dustpol}.} 
\label{fig_pa_hist}
\end{figure}

Since the SMA is more sensitive to small-scale magnetic fields, we computed the uncertainty-weighted averages of SMA polarization angles ($\theta_{SMA}$) to evaluate the averaged orientations of the small-scale magnetic fields of the massive dense cores. 
Figure \ref{fig_pa_hist} shows the distributions of the SMA polarization angles of the cores with the derived $\theta_{SMA}$ listed in Table \ref{phy_table}.
The distributions of the SMA polarization angles have standard deviations of $\sim$ 30--40$^{\circ}$, indicating complex magnetic field structures in the cores.
In the meantime, the JCMT polarization segments at the peaks of the sources ($\theta_{JCMT}$, Table \ref{phy_table}) were derived to represent the large-scale magnetic fields.
To show the difference between the large- and small-scale magnetic fields of the cores, the position angles of $\theta_{SMA}$ and $\theta_{JCMT}$ are plotted in Figures \ref{fig_jcmtsma_south}b and \ref{fig_jcmtsma_north}b.
In Cyg-N53, the angle between $\theta_{SMA}$ and $\theta_{JCMT}$ is 83$^{\circ}$, suggesting that the small-scale magnetic fields are almost perpendicular to the large-scale magnetic fields.
In the other five cores, $\theta_{SMA}$ are aligned within 45$^{\circ}$ to $\theta_{JCMT}$.
In particular, Cyg-N48 and Cyg-N51 have almost parallel $\theta_{SMA}$ and $\theta_{JCMT}$.
In the statistical study of high-mass star-forming regions, \citet{2014Zhang} found that 60\% of the massive dense cores have small-scale magnetic fields aligned within 40$^{\circ}$ to large-scale magnetic fields, and 40\% of the cores have small-scale magnetic fields perfectly perpendicular (80$^{\circ}$--90$^{\circ}$) to large-scale magnetic fields.
Interestingly, despite that the sample of our work is small, we obtain similar conclusion as \citet{2014Zhang}: most massive cores have small- and large-scale magnetic fields aligned within 45$^{\circ}$ and the rest of cores have perpendicular small- and large-scale magnetic fields.

\section{Analysis and Discussion}\label{sec_discuss}

\subsection{Orientations of Cores and Magnetic Fields}
\begin{figure}[b]
\includegraphics[scale=1]{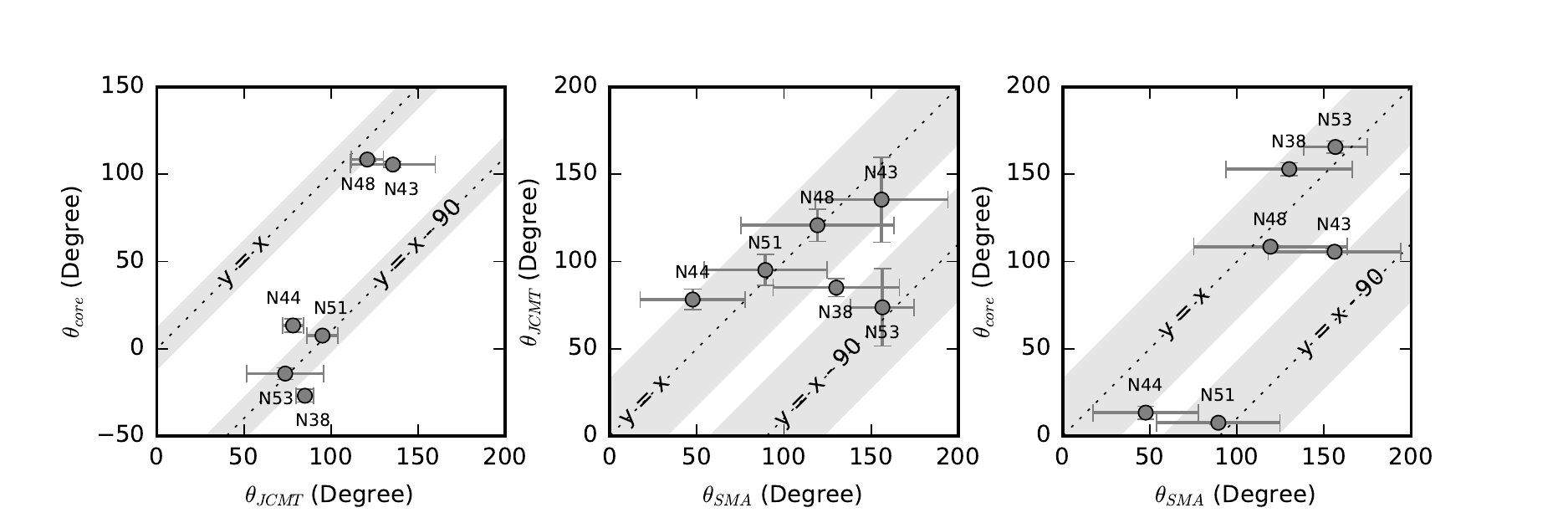}
\caption{Correlations between $\theta_{core}$, $\theta_{JCMT}$, and $\theta_{SMA}$.
The widths of the shadow zones are the means of the error bars in $\theta_{JCMT}$ and $\theta_{SMA}$.
In order to present the correlation, some of the position angles are shifted by 180$\arcdeg$.} 
\label{fig_PA_core_B}
\end{figure}

Figure \ref{fig_PA_core_B} shows the correlations between $\theta_{core}$, $\theta_{JCMT}$, and $\theta_{SMA}$ of the six massive dense cores in the DR21 filament.
$\theta_{core}$ represents the direction of the major axis of cores derived from dendrogram, $\theta_{JCMT}$ represents the orientation of magnetic fields at a 0.1 pc scale, and $\theta_{SMA}$ represents the averaged orientation of magnetic fields at a resolution of 0.03 pc.
In the panel of $\theta_{core}$ versus $\theta_{JCMT}$, the six cores are separated into two groups: one group of parallel alignment and one group of perpendicular alignment, indicating a strong either-parallel-or-perpendicular correlation between $\theta_{core}$ and $\theta_{JCMT}$.
The correlation between $\theta_{JCMT}$ and $\theta_{SMA}$ is weaker: 
while four cores have parallel $\theta_{JCMT}$ and $\theta_{SMA}$ and one core has perpendicular $\theta_{JCMT}$ and $\theta_{SMA}$, Cyg-N38 seems to have $\theta_{JCMT}$ neither parallel nor perpendicular to $\theta_{SMA}$.
Among the three panels, the correlation between $\theta_{SMA}$ and $\theta_{core}$ is much less significant that two of the six cores show no preference in the alignment between $\theta_{SMA}$ and $\theta_{core}$. 

\citet{2014Zhang} found an either-parallel-or-perpendicular alignment between the magnetic fields of cores and the elongation of few-pc-long clumps. The strong correlation between $\theta_{JCMT}$ and $\theta_{core}$ of our data suggests an important role of magnetic fields during the formation of massive dense cores, in agreement with the results of \citet{2014Zhang}. 
However, the weak correlation between $\theta_{SMA}$ and $\theta_{core}$ seems to suggest that the small-scale magnetic fields are not as important as the large-scale magnetic field in supporting a core against gravity.
While the either-parallel-or-perpendicular alignment appears to be a universal property from clouds to cores, why do the alignment
break below the scales of cores?
To explain the break of the either-parallel-or-perpendicular alignment, here we propose a picture that at scales above 0.1 pc, the magnetic fields are strong enough to guide the gas contraction forming structures either aligned or perpendicular to the magnetic fields. When the densities of core scales become large enough to overcome the magnetic fields, gravitational collapse distorts the alignment between the cores and fields, leading to the misalignment below the scales of cores.
In this picture, the misalignment between magnetic fields and cores below 0.1 pc scales could also be the origin of the observed misalignment between outflows and magnetic fields \citep{2014Hull,2014Zhang}. 

\subsection{Derivation of the Magnetic Field Strength} 
\subsubsection{Formalism}
To evaluate the magnetic field strength in the plane of the sky from dust polarization observations,  
the Chandrasekhar--Fermi (hereafter CF; \citeauthor{1953CF} \citeyear{1953CF}) equation is the most widely used method.
The CF equation suggests that the ratio of turbulence to magnetic field strength would lead to a similar level of variation in the magnetic fields as well as in the velocities,
\begin{equation}
\frac{\delta B}{B} \simeq \frac{\delta V_{los}}{V_A}, 
\end{equation}  
where $B$ is the strength of the magnetic field, $\delta B$ is the variation about $B$, $\delta V_{los}$ is the velocity dispersion along the line of sight, and $V_A$ = $B$/$\sqrt{4\pi\rho}$  is the Alfv$\acute{e}$n speed at density $\rho$.

Recently, a statistical modification of the CF equation has been proposed, which avoids inaccurate estimates of field strengths due to a large-scale field model in the CF method \citep{2009Hildebrand} and includes the signal integration across telescope beam and through line-of-sight depth of the source \citep{2009Houde,2011Houde,2016Houde}.    
Assuming that the plane-of-sky component of magnetic fields is composed of a large-scale ordered component $B_0$ and a small-scale turbulent component $B_t$, \citet{2016Houde} suggests that if the turbulence correlation length $\delta$ is much smaller than the thickness of the cloud $\Delta^{\prime}$, 
the ratio of $B_t$ to $B_0$ can be evaluated from the dispersion function of the observed polarization angles (i.e., angular dispersion function),
\begin{multline}
1 - \langle \cos\left[\Delta\Phi\left(l\right)\right]\rangle \simeq \sum_{j=1}^\infty a_{2j}^{\prime}l^{2j} + \left[ \frac{N}{1+N\langle B_0^2\rangle/\langle B_t^2\rangle}\right] \\
 \times\left\{\frac{1}{N_1}\left[1-e^{-l^2/2(\delta^2+2W_1^2)}\right] + \frac{1}{N_2}\left[1-e^{-l^2/2(\delta^2+2W_2^2)}\right]\right. \\
\left. - \frac{2}{N_{12}}\left[1-e^{-l^2/2(\delta^2+W_1^2+W_2^2)}\right] \right\},
\label{eq_adf}
\end{multline} 
where $\Delta\Phi\left(l\right)$ is the difference between polarization angles measured for all pairs of polarization segments separated by a distance $l$, 
the summation is a Taylor expansion representing the structure in the $B_0$ that does not involve turbulence,
$W_1$ and $W_2$ are the widths (i.e., the FWHM beam divided by $\sqrt{8\ln{2}}$) of the synthesis beam and the low-frequency filtering effect, 
and $N$ is the number of turbulent cells along the line of sight obtained by
\begin{equation}
N_1 = \frac{(\delta^2+2W_1^2)\Delta^\prime}{\sqrt{2\pi}\delta^3},
\end{equation}  
\begin{equation}
N_2 = \frac{(\delta^2+2W_2^2)\Delta^\prime}{\sqrt{2\pi}\delta^3},
\end{equation} 
\begin{equation}
N_{12} = \frac{(\delta^2+W_1^2+W_2^2)\Delta^\prime}{\sqrt{2\pi}\delta^3},
\end{equation} 
\begin{equation}
N= \left(\frac{1}{N_1}+\frac{1}{N_2}-\frac{2}{N_{12}}\right)^{-1}.
\end{equation}
The corresponding normalized signal-integrated turbulence autocorrelation function is
\begin{multline}
b^2(l) = \left[ \frac{N}{1+N\langle B_0^2\rangle/\langle B_t^2\rangle}\right]
\left[ \frac{1}{N_1} e^{-l^2/2(\delta^2+2W_1^2)} \right. \\
\left. + \frac{1}{N_2} e^{-l^2/2(\delta^2+2W_2^2)} - \frac{2}{N_{12}} e^{-l^2/2(\delta^2+W_1^2+W_2^2)}\right].
\end{multline} 
Since $B_t$ is the source of perturbation in $B_0$, $B_t$ is identical to the $\delta B$ in the CF equation (Equation 4).   
Therefore, one can apply the $\langle B_t^2\rangle / \langle B_0^2\rangle$ derived from Equation 5 to 
evaluate the magnetic field strength on the plane of sky based on the CF equation
\begin{equation}
\langle B_0^2\rangle^{1/2} = \sqrt{4\pi\rho} \delta V_{los} \left[\frac{\langle B_t^2\rangle}{\langle B_0^2\rangle}\right]^{-1/2}.
\label{eq_cf}
\end{equation}

\subsubsection{Angular Dispersion Function of SMA Polarization Data}
\begin{figure}[b]
\includegraphics[scale=0.9]{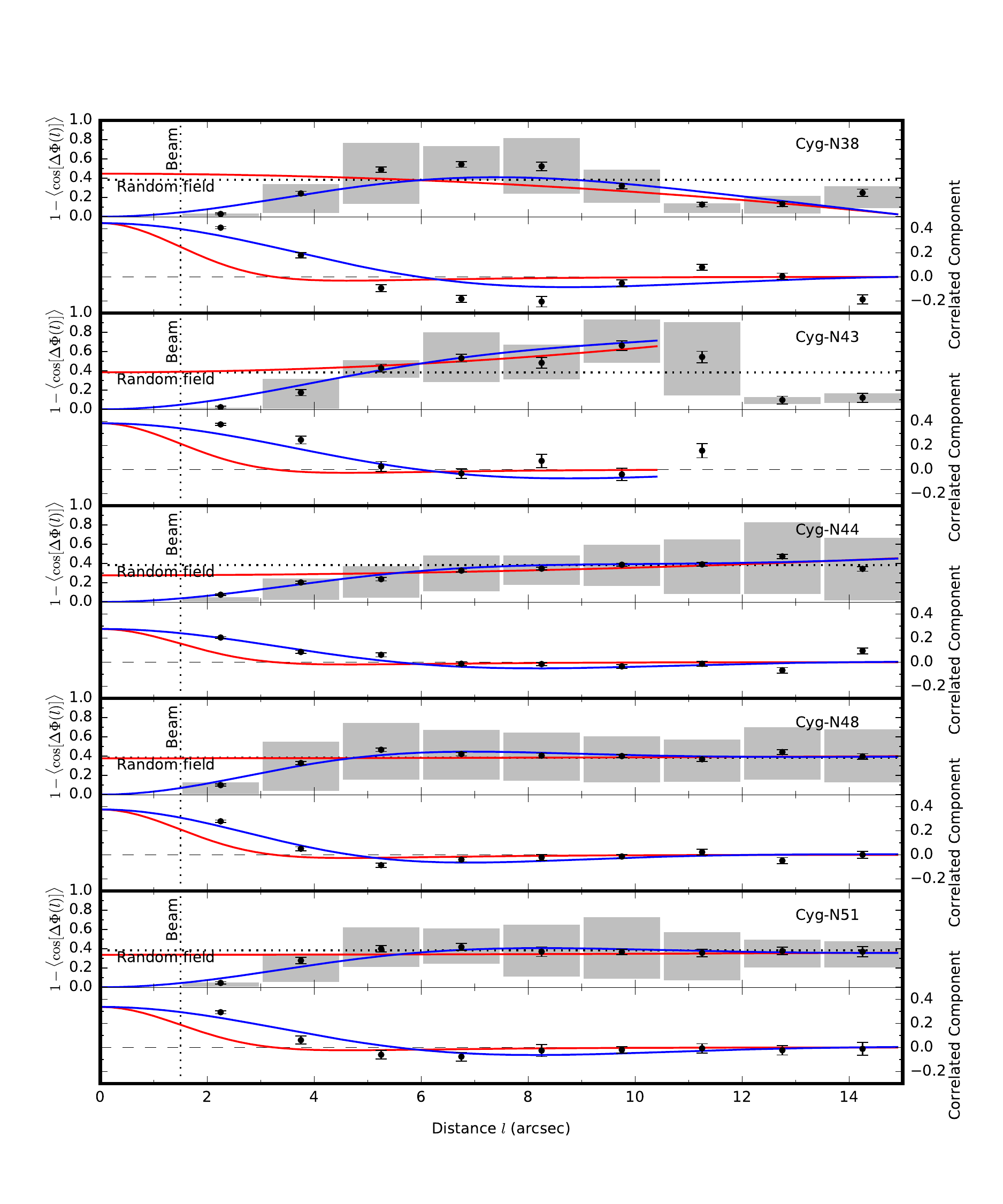}
\caption{Dispersion analysis of the SMA polarization segments toward five cores of the DR21 filament. For each source, the analysis of the angular dispersion function is plotted in the top panel, and the correlated component of the dispersion function is plotted in the bottom panel. 
Top panels: angular dispersion function, $1 - \langle \cos\left[\Delta\Phi\left(l\right)\right]\rangle$, binned in 1$\farcs$5 for a Nyquist sampling of the 3$\arcsec$--4$\arcsec$ beam.
The dots represent the mean values of the data, and the error bars show the standard deviations of the mean values.
The gray boxes display the interquartile ranges of the data.
The blue line shows the best fit to the data (Equation 5), and the red line shows the ordered component $a_2^{\prime}l^2 + b^2(0)$ of the best fit.
The dotted vertical and horizontal lines note the beam size and the expected value for random magnetic fields, respectively. 
Bottom panels: the dots represent the correlated component of the best fit to the data.
The red line shows the correlation due to the beam (Equation 18 in \citeauthor{2016Houde} \citeyear{2016Houde}), and the blue line shows the turbulent component $b^2(l)$ (Equation 10) of the best fit.
The dashed line marks the zero value.}
\label{fig_adf}
\end{figure}

We adopt the statistical analysis of angular dispersion function \citep{2016Houde} to estimate the magnetic field strength in the massive dense cores of the DR21 filaments.
The polarization segments of Cyg-N53 are too few to perform the analysis. 
Figure \ref{fig_adf} shows the angular dispersion functions of the other five cores in the DR21 filament using the SMA polarization segments.
Owing to the limited angular resolution, the angular dispersion function is close to zero when the length scale $l$ is smaller than the beam.
The angular dispersion functions of the cores other than Cyg-N44 smoothly increase with the length scale and reach a maximum at 5$\arcsec$--10$\arcsec$. 
Since the angular dispersion function at small distance is dominated by the turbulent magnetic fields and at large distance dominated by the ordered magnetic fields, the maximum at 5$\arcsec$--10$\arcsec$ indicates a characteristic scale of $\sim$ 0.05 pc (10$^4$ AU) that the magnetic fields change from being turbulent into ordered structures.

At a large distance, the angular dispersion functions of the four cores can be divided into two categories.
In Cyg-N44, Cyg-N48, and Cyg-N51, the angular dispersion functions at scales $\ga$ 8$\arcsec$ are close to random fields ($1-\cos 52^{\arcdeg} = 0.384$; \citeauthor{2010Poidevin} \citeyear{2010Poidevin}), likely representing a 
slow transition of magnetic fields from random structures in the cores to the ordered structure of the filament. 
In Cyg-N38 and Cyg-N43, the maximum of angular dispersion function is higher than the random field value, and the decrease of angular dispersion function is steep with a decline of $\sim$ 0.4 within a distance of 4$\arcsec$. 
Why the the angular dispersion functions of Cyg-N38 and Cyg-N43 are different from those of the other three cores?
Cyg-N38 and Cyg-N43 both may have two groups of polarization segments in different orientations (see Section 3.1). 
The polarization segments within each group contribute low angular dispersions at small and large distances, whereas the angular dispersion measured from one group to the other group contributes a high value at an intermediate distance, resulting in the low-high-low profiles of the angular dispersion functions in Cyg-N38 and Cyg-N43.
In addition to the low-high-low profiles, the distributions of the data also suggest that the angular dispersion functions of Cyg-N38 and Cyg-N43 are resulted from the two groups of magnetic fields rather than random fields.
The gray boxes in Figure \ref{fig_adf} represent the interquartile distributions (25\%--75\%) of the angular dispersion functions.
The angular dispersion functions measured within each group should give a narrower interquartile distribution than those measured from one group to anther. Therefore, the narrower interquartile distributions at small and large distances and the wider interquartile distributions at intermediate distances of Cyg-N38 and Cyg-N43 data are in agreement with the angular dispersion functions from two groups of magnetic fields.

\subsubsection{Derived Magnetic Field Strength}
We use the nonlinear least-squares Marquardt--Levenberg algorithm\footnote{The \texttt{scipy.optimize} package of python} to evaluate the parameters of $\delta$, $\langle B_t^2 \rangle / \langle B_0^2 \rangle$, and $a_{2j}^\prime$ in Equation \ref{eq_adf}, and then use the CF equation (Equation \ref{eq_cf}) to derive the magnetic field strength in the plane of the sky.
For our observations with beams of 3$\arcsec$--4$\arcsec$ and a low-frequency filter of 10$\arcsec$, we use $W_1$ of 1$\farcs$5 and $W_2$ of 4$\farcs$2 in the fitting.
We use the mean FHMW derived from dendrograms as the effective thickness $\Delta^\prime$ of our sources.
We estimate the $\delta V_{lsr}$ from the velocity dispersion of the H$^{13}$CN 4--3 emission in our data (Paper II), since the emission of H$^{13}$CN is well correlated with the dust emission in the massive dense cores of the DR21 filament \citep{2011aCsengeri}.  
The fitting range is from the beam to 15$\arcsec$, and the parameters $a_{2j}^\prime$ are reduced to its first order $a_2^\prime$ because the fitting range is small.
Due to the steep drop in the angular dispersion function of Cyg-N43, the fitting of Cyg-N43 failed to converge unless the maximum fitting distance narrowed down to 10$\arcsec$.
Although the fitting of Cyg-N43 cannot reproduce the angular dispersion function at large distance, the derived magnetic field strength of Cyg-N43 should still be reliable, since the fitting of $\langle B_t^2 \rangle / \langle B_0^2 \rangle$ is mainly regulated by the angular dispersion at small distance and $\langle B_t^2 \rangle / \langle B_0^2 \rangle$ is the only output of the fitting required in the CF equation.

The best fits of angular dispersion functions are shown in Figure \ref{fig_adf}, and the fitted parameters are listed in Table \ref{phy_table2}. 
The correlation lengths $\delta$ of the cores are from 2$\farcs$2 to 3$\farcs$8 (15 to 30 mpc), well resolved by the $W_1$ of 1$\farcs$5. 
The number of turbulent cells $N$ along the line of sight of the sources is about 5--7, and the $\langle B_t^2 \rangle / \langle B_0^2 \rangle$ of the sources is about 3--5, both suggesting that the local turbulent magnetic fields are stronger than the ordered large-scale magnetic fields.
The $\langle B_t^2 \rangle / \langle B_0^2 \rangle$ of the filament is 0.16, significantly smaller than those of the cores. 
Considering that the $B_0$ in the calculations of filament and cores are the same component of the ordered large-scale magnetic field, the increase of $\langle B_t^2 \rangle / \langle B_0^2 \rangle$ indicates an enhancement of the turbulent magnetic fields from the filament to the cores. 

The derived magnetic field strengths on the plane of sky are $\sim$ 0.56 mG in Cyg-N38, $\sim$ 0.42 mG in Cyg-N43, $\sim$ 1.71 mG in Cyg-N44, $\sim$ 0.48 mG in Cyg-N48, and $\sim$ 0.46 mG in Cyg-N51.
Using the same SMA data, the 1.71 mG of Cyg-N44 in this work derived using the dispersion function for interferometry data \citep{2016Houde} is more accurate than the 2.1 mG in \citet{2013Girart} derived using the dispersion function for single-dish data \citep{2009Houde}.
In addition, 1.71 mG is closer than 2.1 mG to the values of 0.9--1.3 mG derived from the BIMA polarization data \citep{2003Lai}, 1.7 mG derived from the velocity dispersion study of the JCMT data \citep{2010Hezareh}, and 1.2 mG derived from the CARMA polarization data \citep{2016Houde}.
For the other four cores, the strengths of 0.4--0.6 mG are close to the strength of the DR21 filament.

To estimate the total magnetic field strength $B_{total}$, we adopt the Zeeman observations at a resolution of 23$\arcsec$ of the CN lines toward Cyg-N44 which probe two line-of-sight magnetic field strengths ($B_{los}$) in two velocity components: $B_{los}$ = -0.36 mG at $v_{LSR}$ = -4.7 km s$^{-1}$ and $B_{los}$ = -0.71 mG at $v_{LSR}$ = -0.9 km s$^{-1}$ \citep{1999Crutcheretal}. Since the -4.7 km s$^{-1}$ component is found to be associated with the Cyg-N44 core and the -0.9 km s$^{-1}$ is the systematic velocity of the filament, the line-of-sight field strengths of Cyg-N44 and the DR21 filament are suggested to be -0.36 mG and -0.71 mG, respectively \citep{2012Crutcher, 2013Girart}.
For the other four cores, we assume an energetic equal partition of magnetic fields ($B_{total} = \sqrt{3/2} B_0$). The derived total magnetic field strengths are listed in Table \ref{phy_table2}. 
Considering the assumption of the energetic equal partition and the dependance of the CF equation (Equation \ref{eq_cf}) to the density which has a 30\%-50\% uncertainty due to the temperature (Section 3.3), we estimate that the uncertainty in the total magnetic field strength is about 50\% of the value.

\begin{figure}[b]
\includegraphics[scale=1]{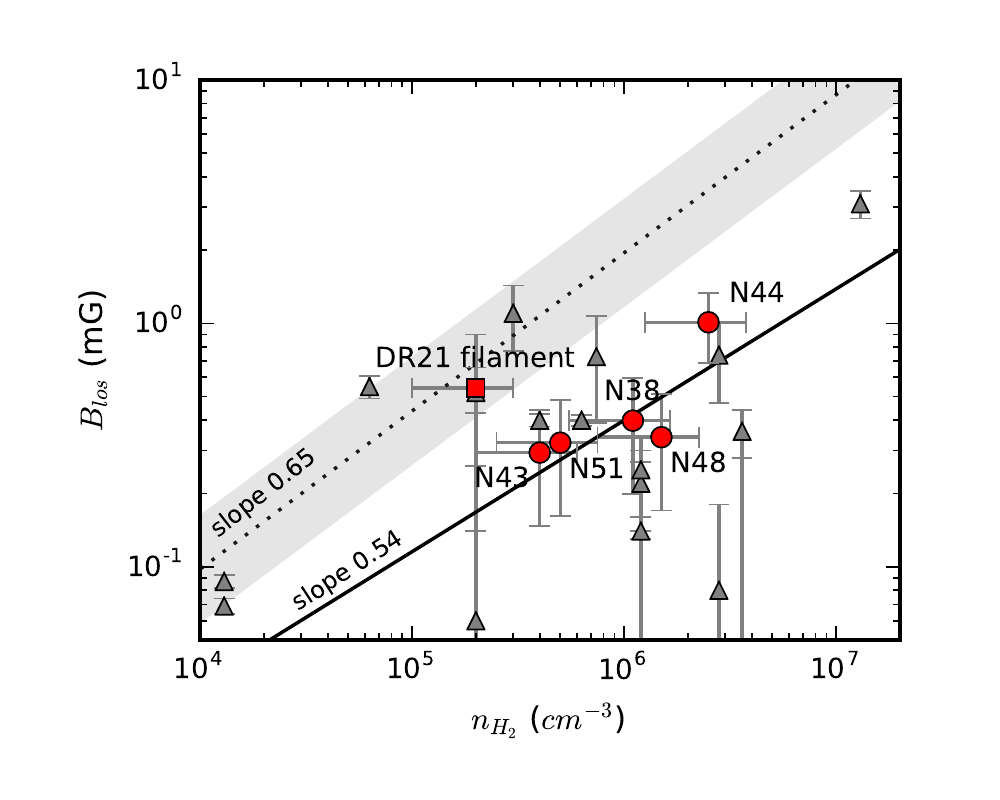}
\caption Line-of-sight magnetic field strength against volume density of molecular cloud. The red circles and red square denote the massive cores and the DR21 filament in this work. The triangles denote the Zeeman observations in \citet{2010Crutcher}. 
Note that despite the Zeeman observations are probing similar density scales, the spatial scales of the Zeeman observations are about 10 times larger than our observations.     
The dotted line represents the model of maximum magnetic field strength scaling with volume density as $B \propto n^{0.65}$ proposed in \citet{2010Crutcher} along with its uncertainty shown as the shadow zone.
The solid line represents the single power-law fit of the massive cores in the DR21 filament.
\label{fig_B_den}
\end{figure}

Figure \ref{fig_B_den} shows the $B_{los}$ versus the volume density of the massive dense cores and the DR21 filament, assuming the most probable value of $B_{los} = B_{total}/\sqrt{3}$.
The magnetic field strengths of the cores are systematically smaller than the strength of the filament, indicating that the true $B_{los}$ of the cores may be stronger than what we assumed. 
Using the data points of the five massive dense cores, a single power-law fit gives a slope of 0.54 $\pm$ 0.30. 
The slope of 0.54 is mainly accounted for the factor of $\sqrt{\rho}$ in the CF equation (Equation \ref{eq_cf}), and the large uncertainty in the fit is resulted from the $\delta V_{los}$ and $\langle B_t^2\rangle / \langle B_0^2\rangle$ which are not scaled with the volume density.
The slope of how magnetic fields scale with gas density offers an important diagnosis: a slope of 2/3 is expected for weak magnetic field model with flux conservation \citep{1966Mestel,2010Crutcher} and 0.44--0.5 is predicted in the ambipolar diffusion model at high density regimes (n$_{H_2}$ $\ga 10^6$ cm$^{-3}$, \citeauthor{1993FM} \citeyear{1993FM}; \citeauthor{2007TM} \citeyear{2007TM}). The recent comprehensive study of Zeeman observations gives a value of 0.65 $\pm$ 0.05 \citep{2010Crutcher}, which is within 1$\sigma$ to our value of 0.54 $\pm$ 0.30. Owing to the $\sqrt{\rho}$ factor in Equation \ref{eq_cf}, the magnetic field strengths derived from the CF equation are not sensitive to study the scaling between magnetic field strength and gas density.

\subsection{Turbulent, Magnetic, and Gravitational Energies}
\subsubsection{$m_s$, $m_A$, $\beta$, and $M/\Phi_B$}
We compute the parameters of the sonic Mach number $m_s$, the Alfv$\acute{e}$n Mach number $m_A$, and the ratio of thermal to magnetic pressures $\beta$ as the following:
\begin{equation}
m_s = \sqrt{3}\sigma_v/c_s, 
\end{equation} 
\begin{equation}
m_A = \sqrt{3}\sigma_v/V_A, 
\end{equation} 
\begin{equation}
\beta=2\left(m_A/m_s\right)^2,
\end{equation} 
where $c_s = \sqrt{\gamma k_BT/\mu m_H}$ is the sound speed at temperature $T$ using adiabatic index $\gamma = 5/3$ and mean molecular weight $\mu = 2.33$,  $\sigma_v = \delta V_{los}$ is the one-dimensional velocity dispersion. 
The derived parameters are listed in Table \ref{phy_table3}.
The DR21 filament has supersonic and sub-Alfv$\acute{e}$nic Mach numbers, revealing that magnetic fields are more important than turbulence in the filament.
The $\beta$ of the filament is less than one, indicating that the magnetic pressure is stronger than the thermal pressure.
Since the linewidth in massive cores includes components of infall, rotation, and turbulence (see 4.3.2), here we adopt the $\sigma_v$ of cores as the broadening of non-thermal motions, instead of turbulence.
The supersonic and trans- to super-Alfv$\acute{e}$nic Mach numbers of massive cores indicate strong non-thermal motions.
The cores typically have a $\beta$ value less than one, suggesting that the magnetic pressure, though weaker than the non-thermal pressure, is stronger than the thermal pressure.
The relation of $m_s>m_A > 1> \beta$ is the same as the results based on the Zeeman observations toward Cyg-N44 \citep{1999Crutcher}.

To estimate the relevance of the magnetic fields with respect to the gravitaty, we derive the mass-to-flux ratio $M/\Phi_B = 7.6 \times10^{-21}\left[N_{H_2}/cm^{-2}\right]\left[B_{total}/\mu G\right]^{-1}$ in units of the critical value $1/2\pi\sqrt{G}$ \citep{1976MS,2004Crutcher}. The derived mass-to-flux ratios of the massive cores and the DR21 filament (see Table \ref{phy_table3}) are about equal or larger than one, suggesting that both in the cores and in the filament, the magnetic fields cannot provide enough support against gravitational collapse. For Cyg-N44, the mass-to-flux ratio of 3.4 is consistent with the values of 2.8--10 estimated from independent methods \citep{1999Crutcher,2010Hezareh}. 
  
\subsubsection{Virial Balance}
The virial theorem is a useful tool for analyzing the effects of kinematics, magnetic fields, and self-gravity in molecular clouds \citep{1956MS, 1991Shu, 2007MO}, which can be written as 
\begin{equation}
\frac{1}{2} \ddot{I} = 2 \left( \mathcal{T}-\mathcal{T}_s \right) + \mathcal{M}+ \mathcal{W}.
\end{equation} 
Here $I$ is a quantity proportional to the inertia of the cloud, and the sign of $\ddot{I}$ determines whether the cloud is expanding ($\ddot{I} > 0$) or contracting ($\ddot{I} < 0$). For a cloud with a fixed mass $M$, the term
\begin{equation}
\mathcal{T} = \frac{3}{2}M\sigma_v^2
\end{equation} 
is the total kinetic energy in the cloud including both thermal and non-thermal (turbulence, rotation, infall, and etc.) components. The term $\mathcal{T}_s$ is the surface kinetic term. The term
\begin{equation}
\mathcal{M} = \frac{1}{2}MV_A^2
\end{equation} 
is the magnetic energy for a cloud with no force from ambient magnetic field. The gravitational term $W$ of a spherical cloud with uniform density and radius $R$ is 
\begin{equation}
 \mathcal{W}=-\frac{3}{5}\frac{GM^2}{R}.
\end{equation} 
The kinetic and gravitational terms of a filamentary cloud are different from those of a spherical cloud \citep{2000FP} in that 
\begin{equation}
\mathcal{T}_f = M\sigma_v^2,
\end{equation} 
\begin{equation}
\mathcal{W}_f =-\frac{GM^2}{L},
\end{equation} 
where $L$ is the length of the filamentary cloud.

The derived ratios of $|\mathcal{T}/\mathcal{W}|$ and $|\mathcal{M}/\mathcal{W}|$ are listed in Table \ref{phy_table3} where we adopt a mass of 15210 $M_{\sun}$ and a length of 4.1 pc for the DR21 filament \citep{2012Hennemann}. 
For the massive dense cores, the typical values of $|\mathcal{T}/\mathcal{W}|$ $\ga$ 1 $\ga$ $|\mathcal{M}/\mathcal{W}|$ reveal a relation of $|\mathcal{T}| \ga |\mathcal{W}| \ga |\mathcal{M}|$, suggesting that kinematic energy is more important than magnetic energy during the formation of massive dense cores.
Although the uncertainty in $B_{total}$ could be as large as 50\% of the value, considering that the $|\mathcal{T}/\mathcal{W}|$ is at least five times larger than the $|\mathcal{M}/\mathcal{W}|$, the relation of $|\mathcal{T}| \ga |\mathcal{M}|$ would not change regardless of the uncertainty in $B_{total}$.
In contrary to the relation of $|\mathcal{T}| \ga |\mathcal{M}|$ of the massive dense cores, the filament has the $|\mathcal{M}/\mathcal{W}|$ significantly larger than the $|\mathcal{T}/\mathcal{W}|$ and therefore $|\mathcal{M}| \ga |\mathcal{T}|$.
The transition from the $|\mathcal{M}| \ga |\mathcal{T}|$ in filament to the $|\mathcal{T}| \ga |\mathcal{M}|$ in cores reveals a decrease of magnetic energy and/or an increase of kinematic energy from the parent filament to the daughter cores.
The variation of the magnetic and kinematic energies from filament to cores is also revealed in the transition from $\langle B_t^2 \rangle / \langle B_0^2 \rangle < 1$ of the filament to the $\langle B_t^2 \rangle / \langle B_0^2 \rangle > 1$ of the cores obtained from the analysis of the angular dispersion functions. 
In star formation models, the diffusion of magnetic flux out of cores is expected through the process of ambipolar diffusion arising from the decoupling of neutral particles and magnetic fields in weakly ionized gas \citep{1987Shu,1991Mouschovias} or reconnection diffusion arising from the magnetic reconnection in turbulent gas \citep{1999LV,2010Santos,2012Lazarian}.
In observations, the presence of turbulent ambipolar diffusion in the massive cores of the DR21 filament is suggested from the different linewidths between ionic and neutral molecular line profiles \citep{2010Hezareh,2014Hezareh}, and the the infall motions, rotation motions, and outflows of the cores \citep{2011aCsengeri,2013Duarte,2013Girart,2014Duarte} might explain the increase of kinematic energy.

The $|(2\mathcal{T+M})/\mathcal{W}|$ listed in Table \ref{phy_table3} represents whether the cloud is expanding ($|(2\mathcal{T+M})/\mathcal{W}| > 1$) or contracting ($|(2\mathcal{T+M})/\mathcal{W}| < 1$) without considering the surface term $\mathcal{T}_s$. The $|(2\mathcal{T+M})/\mathcal{W}|$ of the DR21 filament is less than one, in agreement with the global collapse inferred from molecular line observations \citep{2010Schneider}. However, the $|(2\mathcal{T+M})/\mathcal{W}| \ga 1$ of the massive cores seems to be inconsistent with the signatures of star-forming activities supported by the high level of fragments \citep{2010Bontemps}, infall \citep{2011aCsengeri}, and outflows \citep{2013Duarte,2014Duarte} of the cores. 
How to explain the the discrepancy between the derived virial parameters and the fact that there is star formation in the cores?
The width of the DR21 filament is 0.34 pc \citep{2012Hennemann}, and the mean radius of our six cores is 0.068 pc. Given $\mathcal{W} = -\frac{3}{5} \frac{GM^2}{R}$, the $\mathcal{W}$ changes by a factor of 5 when gas collapses from filament to core. The collapse can transfer the $\mathcal{W}$ in the filament to the $\mathcal{T}$ and $\mathcal{M}$ of cores, resulting in an increase of $|(2\mathcal{T+M})/\mathcal{W}|$ from filament to core.
In addition, a large portion in the $\mathcal{T}$ of core could be in the form of kinematic energy,  instead of turbulent energy.
The velocity dispersions of 1.2--2 km s$^{-1}$ of the Cygnus X massive cores is significantly larger than the virial velocity dispersions $\sqrt{\frac{\alpha MG}{5R}} \sim$ 0.4 km s$^{-1}$ of a 20 M$_{\sun}$ and 0.1 pc core with the geometric factor $\alpha$ equal to unity \citep{1994Williams}. 
Cyg-N48 and Cyg-N53 exhibit infall signatures in the H$^{13}$CO lines \citep{2011aCsengeri}, whereas Cyg-N43, Cyg-N44, and Cyg-N48 show rotation-like structures in our molecular line data (\citeauthor{2013Girart}\ \citeyear{2013Girart}, Paper II). 
The infall and rotation motions could be the origin of the large linewidths in the Cygnus X massive cores.
To estimate the contribution of rotational energy to $\mathcal{T}$, we assume solid-body rotation of core with rotational velocity $\omega$ and rotational energy $\mathcal{T}_r = \frac{2}{5} M R^{2} \omega^{2}$.
Adopting the velocity gradient of $\sim$ 30 km s$^{-1}$ pc$^{-1}$ from our data as $\omega$, the ratio of $\mathcal{T}_r$ to $\mathcal{T}$ is in the order of unity. Since the $\mathcal{T}$ of the core is mostly contributed by kinematic energy, the turbulent energy should be smaller than $\mathcal{W}$.

\begin{figure}[b]
\includegraphics[scale=1]{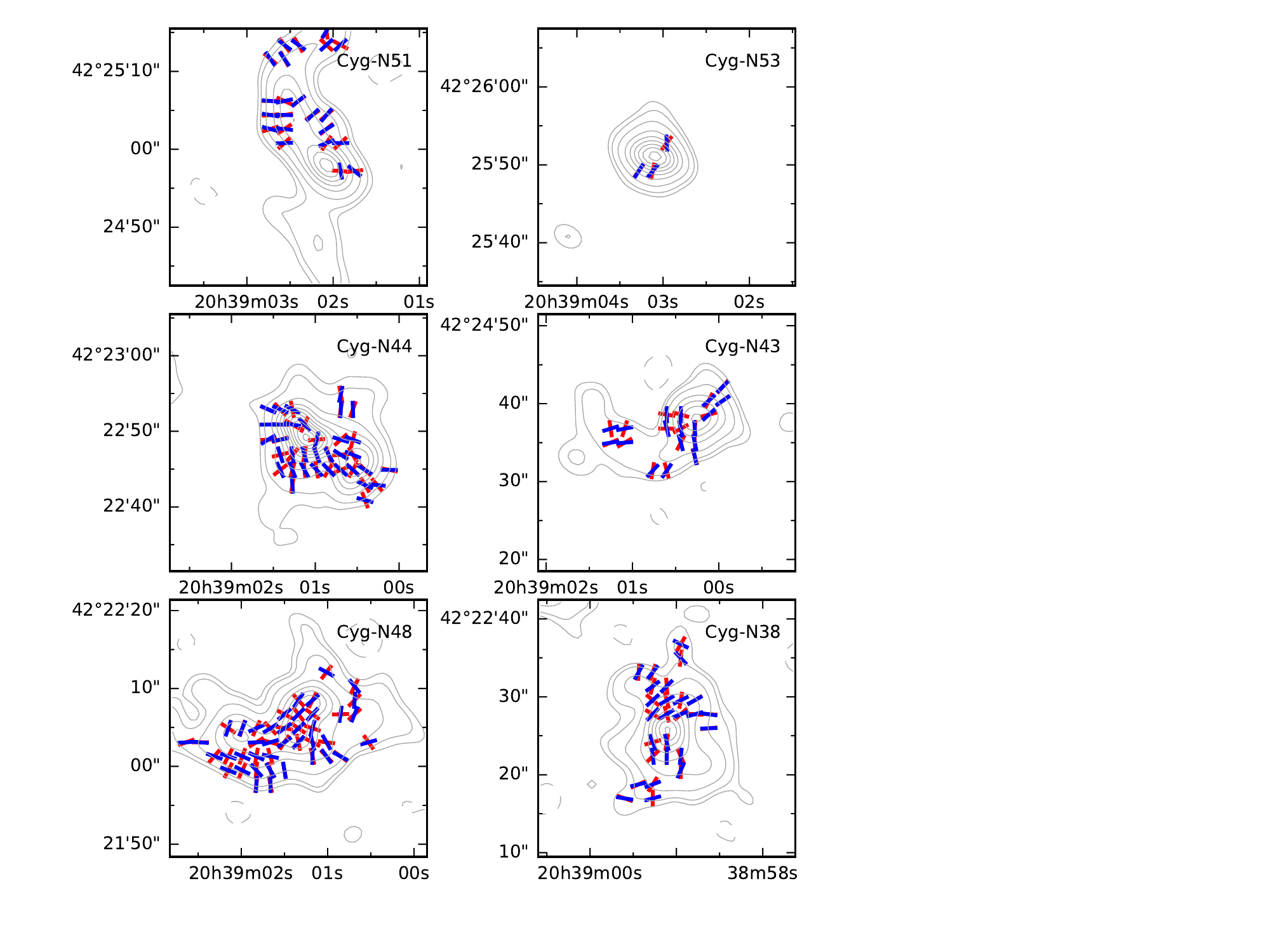}
\caption{The SMA dust polarization maps with magnetic field segments (blue) overlaid with the intensity gradient segments (red) derived from the dust Stokes $I$ continuum contours.}
\label{fig_den_pol}
\end{figure}

\begin{figure}[b]
\includegraphics[scale=1]{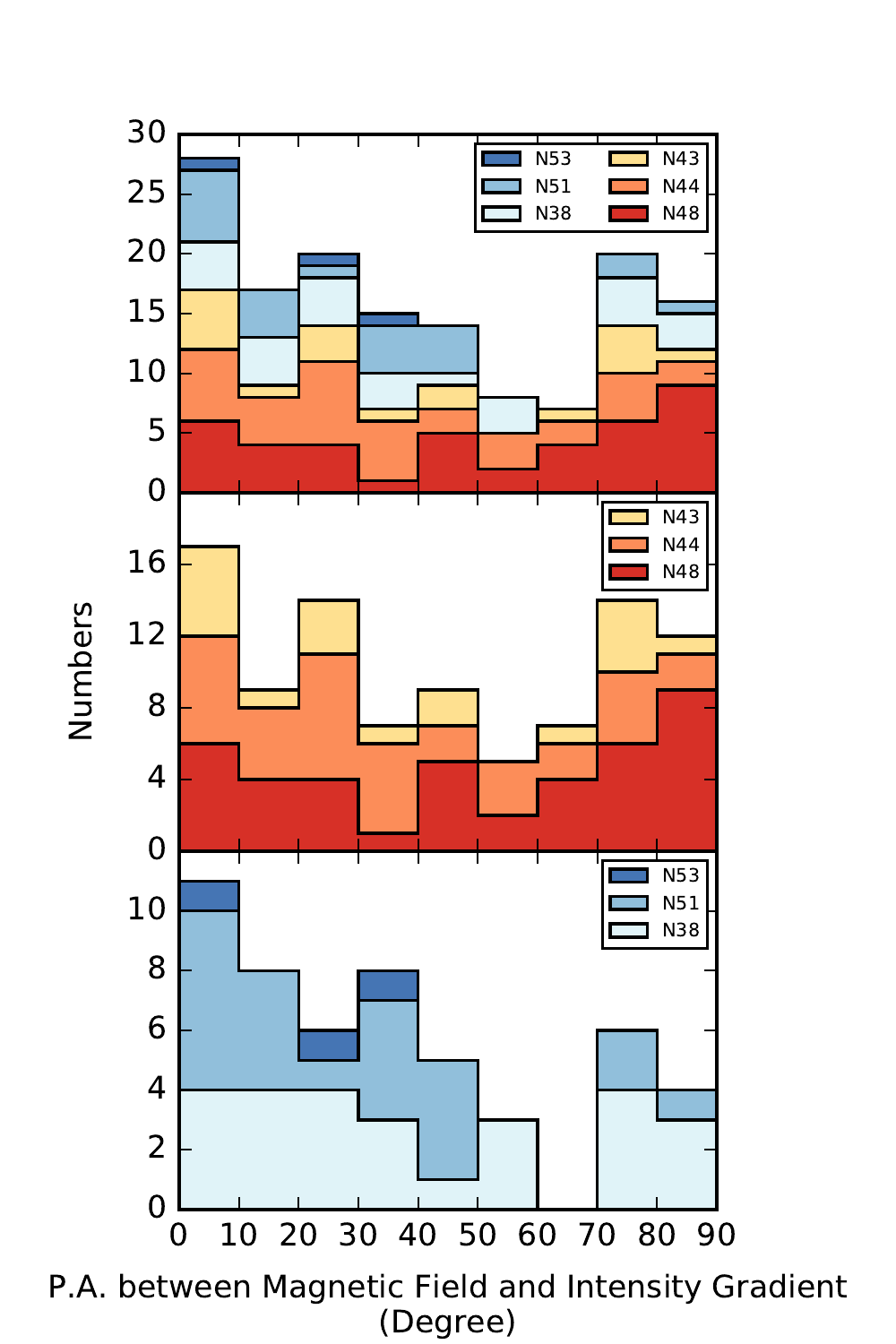}
\caption{Stacked histograms of position angles between the magnetic field segments and intensity gradient segments in Figure \ref{fig_den_pol}. Top: histograms of six massive cores. Middle: histograms of rotation-like cores Cyg-N43, Cyg-N44, and Cyg-N48. Bottom: histograms of non-rotation cores Cyg-N38, Cyg-N51, and Cyg-N53.}
\label{fig_den_pol_hist}
\end{figure}

\begin{figure}[b]
\includegraphics[scale=1]{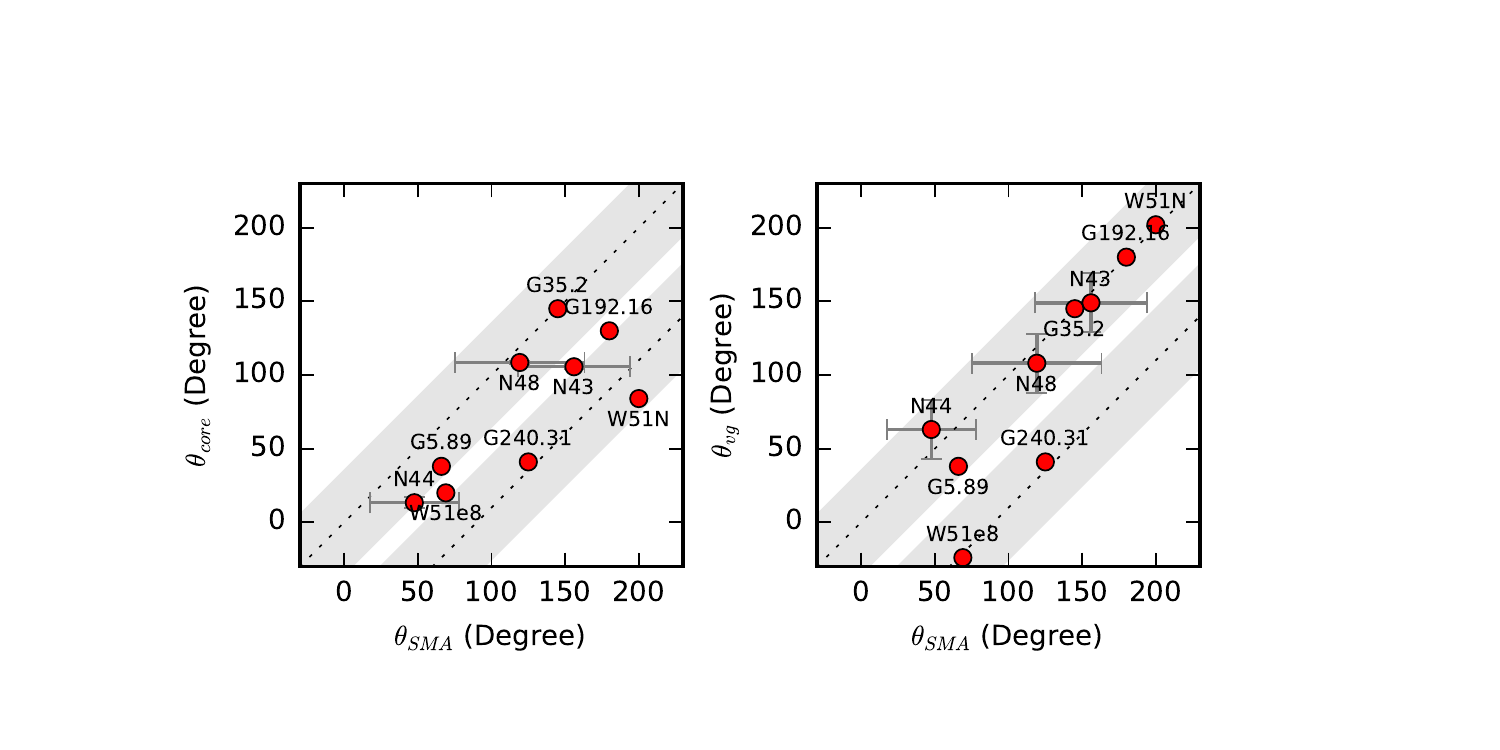}
\caption{Correlations between $\theta_{SMA}$ v.s. $\theta_{core}$ and $\theta_{SMA}$ v.s. $\theta_{vg}$ of the rotation-like cores in the sample of SMA polarization legacy project.
The shadow zones show the mean of the error bars in $\theta_{SMA}$.}
\label{fig_core_sma_vg}
\end{figure}

The large reduction of $\mathcal{W}$ from filament to core may lead to the misalignment between $\theta_{core}$ and $\theta_{SMA}$  through infall and rotation motions of the cores.
The correlation between the magnetic field orientation inferred from dust polarization maps and the intensity gradient of Stokes $I$ dust continuum emission contours has been proposed to evaluate how gravity regulates magnetic fields \citep{2012Koch, 2013Koch}. Based on the assumption that a change in intensity gradient is a measurement for the resulting direction of motion
in the MHD force equation, the magnetic field orientation should be aligned to the intensity gradient when gravity overwhelms magnetic fields, and misalignment between the magnetic field orientation and the intensity gradient is suggested when gravity is relatively less important to magnetic fields. 
Figure \ref{fig_den_pol} displays magnetic field segments and intensity gradient segments of the six massive cores. Figure \ref{fig_den_pol_hist} shows stacked histograms of position angles between magnetic field segments and intensity gradient segments. The rotation-like cores Cyg-N43, Cyg-N44, Cyg-N48 show misalignment between the magnetic field orientation and the intensity gradient, whereas the magnetic fields in the non-rotation cores Cyg-N38, Cyg-N51, Cyg-N53 tend to be aligned with the intensity gradient, suggesting that gravity is important in regulating the orientation of magnetic fields in non-rotation cores but less important in rotation-like cores.
Since the rotational energy of core could be as large as than the gravitation energy, the rotation motion can distort magnetic fields at the core scale, resulting in the misalignment between magnetic fields and the intensity gradient of rotation-like cores. 
Interestingly, the rotation-like cores have $\theta_{SMA}$ parallel to the position angles of velocity gradient ($\theta_{vg}$, Table 3) inferred from our molecular line data. 
Figure \ref{fig_core_sma_vg} shows correlations between $\theta_{SMA}$, $\theta_{core}$, and $\theta_{vg}$ of the sources in the SMA polarization legacy project with clear velocity gradients of rotation-like structures at $\sim$ 0.1 pc scale (G5.89  \citeauthor{2009aTang}\ \citeyear{2009aTang}; W51e8  \citeauthor{1998Zhang}\ \citeyear{1998Zhang}, \citeauthor{2009bTang}\ \citeyear{2009bTang}; 
W51N  \citeauthor{2009Zapata}\ \citeyear{2009Zapata}, \citeauthor{2013Tang}\ \citeyear{2013Tang}; 
G192.12 \citeauthor{2013Liu}\ \citeyear{2013Liu}; G35.2 \citeauthor{2013Qiu} \citeyear{2013Qiu}; G240.31 \citeauthor{2014Qiu}\ \citeyear{2014Qiu}).
The rotation-like cores show a random alignment between $\theta_{SMA}$ and $\theta_{core}$, but strong either-parallel-or-perpendicular correlation between $\theta_{SMA}$ and $\theta_{vg}$, suggesting that rotation plays a more important role than gravity in determining the orientation of magnetic fields.
Overall, the comparisons between magnetic fields and the intensity gradient as well as the correlation between $\theta_{core}$, $\theta_{SMA}$, and $\theta_{vg}$ reveal a picture that magnetic fields of non-rotating core are regulated by gravity, and magnetic fields of rotating-like core are regulated by rotation motion, consistent with the virial paramenter of $|\mathcal{T}| \ga |\mathcal{W}| \ga |\mathcal{M}|$ of the cores.


\section{Conclusions}\label{sec_summary}
We present Submillimeter Array 880 $\mu$m polarization observations of six massive dense cores in the DR21 filament. With the complementary archival polarimetric observations from SCUPOL of the JCMT \citep{2009Matthews}, we are able to characterize the magnetic field properties from the filament to the cores. Our main results are the following:
\begin{enumerate}
\item The SMA dust polarization shows complex magnetic field structures in the massive dense cores, in contrast to the ordered parsec-scale magnetic fields of the filament. Significant depolarization effect is found toward the center of core, and the variation in the SMA polarization angles is $\sim$ 30--40$^{\circ}$ in the cores, again both indicating complex magnetic field structures in the cores.
\item The major axis of core appears to be either parallel or perpendicular to the large-scale magnetic fields inferred from the JCMT polarization map. However, the major axis of the cores shows no preference of alignment to the small-scale magnetic fields inferred from the SMA polarization map. The transition from the either-parallel-or-perpendicular alignment to the random alignment suggests that in the formation of massive dense cores, the magnetic fields below 0.1 pc scales become less important as compared to gravity than the magnetic fields above 0.1 pc scales.
\item Our analysis of angular dispersion functions of SMA polarization segments yields plane-of-sky magnetic field strengths of 1.7 mG in Cyg-N44 and about 0.5 mG in other four cores. Most of the cores have the angular dispersion functions similar to the dispersion of random magnetic field. The derived number of turbulent cells and the derived ratio of turbulent magnetic fields to ordered magnetic fields are both much greater than unity, indicating that the magnetic fields in the cores are dominated by turbulent components.
\item The virial parameters show that in the filament, the gravitational energy $\mathcal{W}$ is dominant over magnetic $\mathcal{M}$ and kinematic $\mathcal{T}$ energies, and the gravitational collapse can transfer the $\mathcal{W}$ of the filament to the $\mathcal{T}$ and $\mathcal{M}$ of the cores, resulting in the relations of $|\mathcal{T}| \ga |\mathcal{W}| \ga |\mathcal{M}|$ and $|(2\mathcal{T+M})/\mathcal{W}| \ga 1$ in the cores.
The virial parameters reveal an overall picture of weakly magnetized cores in a strongly magnetized filament, suggesting that the kinematics arising from gravitational collapse must become more important than magnetic fields during the evolution from filament to massive dense cores.
\end{enumerate}

\acknowledgments
Part of the data were obtained in the SMA legacy project: Filaments, Magnetic Fields, and Massive Star Formation (PI: Qizhou Zhang).
T. C. C. acknowledges the support of the Smithsonian Predoctoral Fellowship and the University Consortium of ALMA-Taiwan (UCAT) Graduate Fellowship.
T. C. C. and S. P. L. are thankful for the support of the Ministry of Science and Technology (MoST) of Taiwan through Grants 
102-2119-M-007-004-MY3, 104-2119-M-007-021, 105-2119-M-007 -022 -MY3, and 105-2119-M-007 -024.
Q. Z. acknowledges the support of the SI Scholarly Studies Awards ``Are Magnetic Fields Dynamically Important in Massive Star Formation?''
J. M. G. acknowledges the support from the MICINN (Spain) AYA2014-57369- C3 grant and the MECD (Spain) PRX15/00435 travel grant. 
K. Q. acknowledges the support from National Natural Science Foundation of China (NSFC) through grants NSFC 11473011 and NSFC 11590781.


\begin{thebibliography}{}
\expandafter\ifx\csname natexlab\endcsname\relax\def\natexlab#1{#1}\fi

\bibitem[{Andr{\'e} {et~al.}(2014)Andr{\'e}, Di~Francesco, Ward-Thompson,
  Inutsuka, Pudritz, \& Pineda}]{2014Andre}
Andr{\'e}, P., Di~Francesco, J., Ward-Thompson, D., {et~al.} 2014, Protostars
  and Planets VI, 27

\bibitem[{Argon {et~al.}(2000)Argon, Reid, \& Menten}]{2000Argon}
Argon, A.~L., Reid, M.~J., \& Menten, K.~M. 2000, The ApJS, 129, 159

\bibitem[{{Bontemps} {et~al.}(2010){Bontemps}, {Motte}, {Csengeri}, \&
  {Schneider}}]{2010Bontemps}
{Bontemps}, S., {Motte}, F., {Csengeri}, T., \& {Schneider}, N. 2010, \aap,
  524, A18

\bibitem[{{Braz} \& {Epchtein}(1983)}]{1983BE}
{Braz}, M.~A., \& {Epchtein}, N. 1983, \aaps, 54, 167

\bibitem[{{Chandrasekhar} \& {Fermi}(1953)}]{1953CF}
{Chandrasekhar}, S., \& {Fermi}, E. 1953, \apj, 118, 113

\bibitem[Chen et al.(2016)]{2016ApJ...829...84C} Chen, C.-Y., King, P.~K., \& Li, Z.-Y.\ 2016, \apj, 829, 84 

\bibitem[{Cortes {et~al.}(2008)Cortes, Crutcher, Shepherd, \&
  Bronfman}]{2008Cortes}
Cortes, P.~C., Crutcher, R.~M., Shepherd, D.~S., \& Bronfman, L. 2008, \apj, 676, 464

\bibitem[{Cortes {et~al.}(2016)Cortes, Girart, Hull, Sridharan, Louvet,
  Plambeck, Li, Crutcher, \& Lai}]{2016Cortes}
Cortes, P.~C., Girart, J.~M., Hull, C. L.~H., {et~al.} 2016, ApJL, 825, L15

\bibitem[{Crutcher(1999)}]{1999Crutcher}
Crutcher, R.~M. 1999, \apj, 520, 706

\bibitem[{{Crutcher}(2004)}]{2004Crutcher}
{Crutcher}, R.~M. 2004, \apss, 292, 225

\bibitem[{Crutcher(2012)}]{2012Crutcher}
Crutcher, R.~M. 2012, ARA\&A, 50, 29

\bibitem[{Crutcher {et~al.}(1999)Crutcher, Troland, Lazareff, Paubert, \&
  Kaz{\`e}s}]{1999Crutcheretal}
Crutcher, R.~M., Troland, T.~H., Lazareff, B., Paubert, G., \& Kaz{\`e}s, I.
  1999, \apj, 514, L121

\bibitem[{{Crutcher} {et~al.}(2010){Crutcher}, {Wandelt}, {Heiles},
  {Falgarone}, \& {Troland}}]{2010Crutcher}
{Crutcher}, R.~M., {Wandelt}, B., {Heiles}, C., {Falgarone}, E., \& {Troland},
  T.~H. 2010, \apj, 725, 466

\bibitem[{{Csengeri} {et~al.}(2011{\natexlab{a}}){Csengeri}, {Bontemps},
  {Schneider}, {Motte}, \& {Dib}}]{2011aCsengeri}
{Csengeri}, T., {Bontemps}, S., {Schneider}, N., {Motte}, F., \& {Dib}, S.
  2011{\natexlab{a}}, \aap, 527, A135

\bibitem[{{Csengeri} {et~al.}(2011{\natexlab{b}}){Csengeri}, {Bontemps},
  {Schneider}, {Motte}, {Gueth}, \& {Hora}}]{2011bCsengeri}
{Csengeri}, T., {Bontemps}, S., {Schneider}, N., {et~al.} 2011{\natexlab{b}},
  \apjl, 740, L5

\bibitem[{{Davis} {et~al.}(2007){Davis}, {Kumar}, {Sandell}, {Froebrich},
  {Smith}, \& {Currie}}]{2007Davis}
{Davis}, C.~J., {Kumar}, M.~S.~N., {Sandell}, G., {et~al.} 2007, \mnras, 374,
  29

\bibitem[{{Di Francesco} {et~al.}(2008){Di Francesco}, {Johnstone}, {Kirk},
  {MacKenzie}, \& {Ledwosinska}}]{2008Di}
{Di Francesco}, J., {Johnstone}, D., {Kirk}, H., {MacKenzie}, T., \&
  {Ledwosinska}, E. 2008, \apjs, 175, 277

\bibitem[{Downes \& Rinehart(1966)}]{1966DR}
Downes, D., \& Rinehart, R. 1966, ApJ, 144, 937

\bibitem[{{Duarte-Cabral} {et~al.}(2014){Duarte-Cabral}, {Bontemps}, {Motte},
  {Gusdorf}, {Csengeri}, {Schneider}, \& {Louvet}}]{2014Duarte}
{Duarte-Cabral}, A., {Bontemps}, S., {Motte}, F., {et~al.} 2014, \aap, 570, A1

\bibitem[{{Duarte-Cabral} {et~al.}(2013){Duarte-Cabral}, {Bontemps}, {Motte},
  {Hennemann}, {Schneider}, \& {Andr{\'e}}}]{2013Duarte}
---. 2013, \aap, 558, A125

\bibitem[{Falgarone {et~al.}(2008)Falgarone, Troland, Crutcher, \&
  Paubert}]{2008Falgarone}
Falgarone, E., Troland, T.~H., Crutcher, R.~M., \& Paubert, G. 2008, \aap, 487, 247

\bibitem[{Fiedler \& Mouschovias(1993)}]{1993FM}
Fiedler, R.~A., \& Mouschovias, T.~C. 1993, ApJ v.415, 415,
  680

\bibitem[{Fiege \& Pudritz(2000)}]{2000FP}
Fiege, J.~D., \& Pudritz, R.~E. 2000, MNRAS, 311, 85

\bibitem[{{Frau} {et~al.}(2014){Frau}, {Girart}, {Zhang}, \& {Rao}}]{2014Frau}
{Frau}, P., {Girart}, J.~M., {Zhang}, Q., \& {Rao}, R. 2014, \aap, 567, A116

\bibitem[{{Galv{\'a}n-Madrid} {et~al.}(2010){Galv{\'a}n-Madrid}, {Zhang},
  {Keto}, {Ho}, {Zapata}, {Rodr{\'{\i}}guez}, {Pineda}, \&
  {V{\'a}zquez-Semadeni}}]{2010GM}
{Galv{\'a}n-Madrid}, R., {Zhang}, Q., {Keto}, E., {et~al.} 2010, \apj, 725, 17

\bibitem[{{Girart} {et~al.}(2009){Girart}, {Beltr{\'a}n}, {Zhang}, {Rao}, \&
  {Estalella}}]{2009Girart}
{Girart}, J.~M., {Beltr{\'a}n}, M.~T., {Zhang}, Q., {Rao}, R., \& {Estalella},
  R. 2009, Science, 324, 1408

\bibitem[{{Girart} {et~al.}(2013){Girart}, {Frau}, {Zhang}, {Koch}, {Qiu},
  {Tang}, {Lai}, \& {Ho}}]{2013Girart}
{Girart}, J.~M., {Frau}, P., {Zhang}, Q., {et~al.} 2013, \apj, 772, 69

\bibitem[{Glenn {et~al.}(1999)Glenn, Walker, \& Young}]{1999Glenn}
Glenn, J., Walker, C.~K., \& Young, E.~T. 1999, \apj, 511,
  812

\bibitem[{{Goodman} {et~al.}(2009){Goodman}, {Rosolowsky}, {Borkin}, {Foster},
  {Halle}, {Kauffmann}, \& {Pineda}}]{2009Goodman}
{Goodman}, A.~A., {Rosolowsky}, E.~W., {Borkin}, M.~A., {et~al.} 2009, \nat,
  457, 63

\bibitem[Hennemann et al.(2012)]{2012A&A...543L...3H} Hennemann, M., Motte, F., Schneider, N., et al.\ 2012, \aap, 543, L3 

\bibitem[{{Hezareh} {et~al.}(2014){Hezareh}, {Csengeri}, {Houde}, {Herpin}, \&
  {Bontemps}}]{2014Hezareh}
{Hezareh}, T., {Csengeri}, T., {Houde}, M., {Herpin}, F., \& {Bontemps}, S.
  2014, \mnras, 438, 663

\bibitem[{Hezareh {et~al.}(2010)Hezareh, Houde, McCoey, \& Li}]{2010Hezareh}
Hezareh, T., Houde, M., McCoey, C., \& Li, H.-b. 2010, ApJ, 720, 603

\bibitem[{{Hildebrand} {et~al.}(2009){Hildebrand}, {Kirby}, {Dotson}, {Houde},
  \& {Vaillancourt}}]{2009Hildebrand}
{Hildebrand}, R.~H., {Kirby}, L., {Dotson}, J.~L., {Houde}, M., \&
  {Vaillancourt}, J.~E. 2009, \apj, 696, 567

\bibitem[{{Ho} {et~al.}(2004){Ho}, {Moran}, \& {Lo}}]{2004Ho}
{Ho}, P.~T.~P., {Moran}, J.~M., \& {Lo}, K.~Y. 2004, \apjl, 616, L1

\bibitem[{{Houde} {et~al.}(2016){Houde}, {Hull}, {Plambeck}, {Vaillancourt}, \&
  {Hildebrand}}]{2016Houde}
{Houde}, M., {Hull}, C.~L.~H., {Plambeck}, R.~L., {Vaillancourt}, J.~E., \&
  {Hildebrand}, R.~H. 2016, \apj, 820, 38

\bibitem[{{Houde} {et~al.}(2011){Houde}, {Rao}, {Vaillancourt}, \&
  {Hildebrand}}]{2011Houde}
{Houde}, M., {Rao}, R., {Vaillancourt}, J.~E., \& {Hildebrand}, R.~H. 2011,
  \apj, 733, 109

\bibitem[{{Houde} {et~al.}(2009){Houde}, {Vaillancourt}, {Hildebrand},
  {Chitsazzadeh}, \& {Kirby}}]{2009Houde}
{Houde}, M., {Vaillancourt}, J.~E., {Hildebrand}, R.~H., {Chitsazzadeh}, S., \&
  {Kirby}, L. 2009, \apj, 706, 1504

\bibitem[{Hull {et~al.}(2014)Hull, Plambeck, Kwon, Bower, Carpenter, Crutcher,
  Fiege, Franzmann, Hakobian, Heiles, Houde, Hughes, Lamb, Looney, Marrone,
  Matthews, Pillai, Pound, Rahman, Sandell, Stephens, Tobin, Vaillancourt,
  Volgenau, \& Wright}]{2014Hull}
Hull, C. L.~H., Plambeck, R.~L., Kwon, W., {et~al.} 2014, ApJS, 213, 13

\bibitem[{{Kirby}(2009)}]{2009Kirby}
{Kirby}, L. 2009, \apj, 694, 1056

\bibitem[{{Koch} {et~al.}(2012){Koch}, {Tang}, \& {Ho}}]{2012Koch}
{Koch}, P.~M., {Tang}, Y.-W., \& {Ho}, P.~T.~P. 2012, \apj, 747, 79

\bibitem[{{Koch} {et~al.}(2013){Koch}, {Tang}, \& {Ho}}]{2013Koch}
---. 2013, \apj, 775, 77

\bibitem[{Koch {et~al.}(2014)Koch, Tang, Ho, Zhang, Girart, Chen, Frau, Li, Li,
  Liu, Padovani, Qiu, Yen, Chen, Ching, Lai, \& Rao}]{2014Koch}
Koch, P.~M., Tang, Y.-W., Ho, P. T.~P., {et~al.} 2014, ApJ, 797, 99

\bibitem[{Krumholz \& Tan(2007)}]{2007KT}
Krumholz, M.~R., \& Tan, J.~C. 2007, \apj, 654, 304

\bibitem[{{Lai} {et~al.}(2001){Lai}, {Crutcher}, {Girart}, \& {Rao}}]{2001Lai}
{Lai}, S.-P., {Crutcher}, R.~M., {Girart}, J.~M., \& {Rao}, R. 2001, \apj, 561,
  864

\bibitem[{{Lai} {et~al.}(2002){Lai}, {Crutcher}, {Girart}, \& {Rao}}]{2002Lai}
---. 2002, \apj, 566, 925

\bibitem[{{Lai} {et~al.}(2003){Lai}, {Girart}, \& {Crutcher}}]{2003Lai}
{Lai}, S.-P., {Girart}, J.~M., \& {Crutcher}, R.~M. 2003, \apj, 598, 392

\bibitem[Lazarian et al.(2012)]{2012Lazarian} Lazarian, A., Esquivel, A., \& Crutcher, R.\ 2012, \apj, 757, 154 

\bibitem[Lazarian \& Vishniac(1999)]{1999LV} Lazarian, A., \& Vishniac, E.~T.\ 1999, \apj, 517, 700

\bibitem[{Li {et~al.}(2013)Li, Fang, Henning, \& Kainulainen}]{2013Li}
Li, H.-b., Fang, M., Henning, T., \& Kainulainen, J. 2013, MNRAS, 436, 3707

\bibitem[{Li {et~al.}(2014)Li, Goodman, Sridharan, Houde, Li, Novak, \&
  Tang}]{2014Li}
Li, H.-b., Goodman, A., Sridharan, T.~K., {et~al.} 2014, Protostars and Planets
  VI, 101

\bibitem[{Li {et~al.}(2015)Li, Yuen, Otto, Leung, Sridharan, Zhang, Liu, Tang,
  \& Qiu}]{2015Li}
Li, H.-b., Yuen, K.~H., Otto, F., {et~al.} 2015, Nature, 520, 518

\bibitem[{{Liu} {et~al.}(2012){Liu}, {Jim{\'e}nez-Serra}, {Ho}, {Chen},
  {Zhang}, \& {Li}}]{2012Liu}
{Liu}, H.~B., {Jim{\'e}nez-Serra}, I., {Ho}, P.~T.~P., {et~al.} 2012, \apj,
  756, 10

\bibitem[{{Liu} {et~al.}(2013){Liu}, {Qiu}, {Zhang}, {Girart}, \&
  {Ho}}]{2013Liu}
{Liu}, H.~B., {Qiu}, K., {Zhang}, Q., {Girart}, J.~M., \& {Ho}, P.~T.~P. 2013,
  \apj, 771, 71

\bibitem[Li et al.(2016)]{2016Li} Li, P.~S., Klein, R.~I., \& McKee, C.~F.\ 2016, From Interstellar Clouds to Star-Forming Galaxies: Universal Processes, 315, 103

\bibitem[{{Marrone}(2006)}]{2006Marrone}
{Marrone}, D.~P. 2006, PhD thesis, Harvard University

\bibitem[{Marrone \& Rao(2008)}]{2008MR}
Marrone, D.~P., \& Rao, R. 2008, Millimeter and Submillimeter Detectors and
  Instrumentation for Astronomy IV. Edited by Duncan, 7020, 70202B

\bibitem[{Matthews {et~al.}(2009)Matthews, McPhee, Fissel, \&
  Curran}]{2009Matthews}
Matthews, B.~C., McPhee, C.~A., Fissel, L.~M., \& Curran, R.~L. 2009, ApJS, 182, 143

\bibitem[{Mayer {et~al.}(1973)Mayer, Waak, Cheung, \& Chui}]{1973Mayer}
Mayer, C.~H., Waak, J.~A., Cheung, A.~C., \& Chui, M.~F. 1973, ApJ, 182, L65

\bibitem[{{McKee} \& {Ostriker}(2007)}]{2007MO}
{McKee}, C.~F., \& {Ostriker}, E.~C. 2007, \araa, 45, 565

\bibitem[{Mestel(1966)}]{1966Mestel}
Mestel, L. 1966, MNRAS, 133, 265

\bibitem[{Mestel \& Spitzer(1956)}]{1956MS}
Mestel, L., \& Spitzer, L.~J. 1956, MNRAS, 116, 503

\bibitem[{{Minchin} \& {Murray}(1994)}]{1994MM}
{Minchin}, N.~R., \& {Murray}, A.~G. 1994, \aap, 286, 579

\bibitem[{{Motte} {et~al.}(2007){Motte}, {Bontemps}, {Schilke}, {Schneider},
  {Menten}, \& {Brogui{\`e}re}}]{2007Motte}
{Motte}, F., {Bontemps}, S., {Schilke}, P., {et~al.} 2007, \aap, 476, 1243

\bibitem[{Mouschovias \& Spitzer(1976)}]{1976MS}
Mouschovias, T.~C., \& Spitzer, L.~J. 1976, ApJ, 210, 326

\bibitem[{Mouschovias(1991)}]{1991Mouschovias}Mouschovias, T.~C. 1991, ApJ, 373, 169

\bibitem[{Naghizadeh-Khouei \& Clarke(1993)}]{1993NC}
Naghizadeh-Khouei, J., \& Clarke, D. 1993, \aap, 274,
  968

\bibitem[{Nakamura \& Li(2008)}]{2008NL}
Nakamura, F., \& Li, Z.-Y. 2008, \apj, 687, 354

\bibitem[{{Ossenkopf} \& {Henning}(1994)}]{1994OH}
{Ossenkopf}, V., \& {Henning}, T. 1994, \aap, 291, 943


\bibitem[{Palau {et~al.}(2013)Palau, Fuente, Girart, Estalella, Ho,
  Sanchez-Monge, Fontani, Busquet, Commer{\c c}on, Hennebelle, Boissier, Zhang,
  Cesaroni, \& Zapata}]{2013Palau}
Palau, A., Fuente, A., Girart, J.~M., {et~al.} 2013, \apj,
  762, 120

\bibitem[{{Peretto} {et~al.}(2013){Peretto}, {Fuller}, {Duarte-Cabral},
  {Avison}, {Hennebelle}, {Pineda}, {Andr{\'e}}, {Bontemps}, {Motte},
  {Schneider}, \& {Molinari}}]{2013Peretto}
{Peretto}, N., {Fuller}, G.~A., {Duarte-Cabral}, A., {et~al.} 2013, \aap, 555,
  A112

\bibitem[{Pestalozzi {et~al.}(2005)Pestalozzi, Minier, \&
  Booth}]{2005Pestalozzi}
Pestalozzi, M.~R., Minier, V., \& Booth, R.~S. 2005, \aap, 432, 737

\bibitem[{{Planck Collaboration} {et~al.}(2016){Planck Collaboration}, {Ade},
  {Aghanim}, {Alves}, {Arnaud}, {Arzoumanian}, {Ashdown}, {Aumont},
  {Baccigalupi}, {Banday}, {Barreiro}, {Bartolo}, {Battaner}, {Benabed},
  {Beno{\^i}t}, {Benoit-L{\'e}vy}, {Bernard}, {Bersanelli}, {Bielewicz},
  {Bock}, {Bonavera}, {Bond}, {Borrill}, {Bouchet}, {Boulanger}, {Bracco},
  {Burigana}, {Calabrese}, {Cardoso}, {Catalano}, {Chiang}, {Christensen},
  {Colombo}, {Combet}, {Couchot}, {Crill}, {Curto}, {Cuttaia}, {Danese},
  {Davies}, {Davis}, {de Bernardis}, {de Rosa}, {de Zotti}, {Delabrouille},
  {Dickinson}, {Diego}, {Dole}, {Donzelli}, {Dor{\'e}}, {Douspis}, {Ducout},
  {Dupac}, {Efstathiou}, {Elsner}, {En{\ss}lin}, {Eriksen}, {Falceta-Gon{\c
  c}alves}, {Falgarone}, {Ferri{\`e}re}, {Finelli}, {Forni}, {Frailis},
  {Fraisse}, {Franceschi}, {Frejsel}, {Galeotta}, {Galli}, {Ganga}, {Ghosh},
  {Giard}, {Gjerl{\o}w}, {Gonz{\'a}lez-Nuevo}, {G{\'o}rski}, {Gregorio},
  {Gruppuso}, {Gudmundsson}, {Guillet}, {Harrison}, {Helou}, {Hennebelle},
  {Henrot-Versill{\'e}}, {Hern{\'a}ndez-Monteagudo}, {Herranz}, {Hildebrandt},
  {Hivon}, {Holmes}, {Hornstrup}, {Huffenberger}, {Hurier}, {Jaffe}, {Jaffe},
  {Jones}, {Juvela}, {Keih{\"a}nen}, {Keskitalo}, {Kisner}, {Knoche}, {Kunz},
  {Kurki-Suonio}, {Lagache}, {Lamarre}, {Lasenby}, {Lattanzi}, {Lawrence},
  {Leonardi}, {Levrier}, {Liguori}, {Lilje}, {Linden-V{\o}rnle},
  {L{\'o}pez-Caniego}, {Lubin}, {Mac{\'{\i}}as-P{\'e}rez}, {Maino},
  {Mandolesi}, {Mangilli}, {Maris}, {Martin}, {Mart{\'{\i}}nez-Gonz{\'a}lez},
  {Masi}, {Matarrese}, {Melchiorri}, {Mendes}, {Mennella}, {Migliaccio},
  {Miville-Desch{\^e}nes}, {Moneti}, {Montier}, {Morgante}, {Mortlock},
  {Munshi}, {Murphy}, {Naselsky}, {Nati}, {Netterfield}, {Noviello}, {Novikov},
  {Novikov}, {Oppermann}, {Oxborrow}, {Pagano}, {Pajot}, {Paladini},
  {Paoletti}, {Pasian}, {Perotto}, {Pettorino}, {Piacentini}, {Piat},
  {Pierpaoli}, {Pietrobon}, {Plaszczynski}, {Pointecouteau}, {Polenta},
  {Ponthieu}, {Pratt}, {Prunet}, {Puget}, {Rachen}, {Reinecke}, {Remazeilles},
  {Renault}, {Renzi}, {Ristorcelli}, {Rocha}, {Rossetti}, {Roudier},
  {Rubi{\~n}o-Mart{\'{\i}}n}, {Rusholme}, {Sandri}, {Santos}, {Savelainen},
  {Savini}, {Scott}, {Soler}, {Stolyarov}, {Sudiwala}, {Sutton}, {Suur-Uski},
  {Sygnet}, {Tauber}, {Terenzi}, {Toffolatti}, {Tomasi}, {Tristram}, {Tucci},
  {Umana}, {Valenziano}, {Valiviita}, {Van Tent}, {Vielva}, {Villa}, {Wade},
  {Wandelt}, {Wehus}, {Ysard}, {Yvon}, \& {Zonca}}]{2016Planck}
{Planck Collaboration}, {Ade}, P.~A.~R., {Aghanim}, N., {et~al.} 2016, \aap,
  586, A138

\bibitem[{{Poidevin} {et~al.}(2010){Poidevin}, {Bastien}, \&
  {Matthews}}]{2010Poidevin}
{Poidevin}, F., {Bastien}, P., \& {Matthews}, B.~C. 2010, \apj, 716, 893

\bibitem[{{Press} {et~al.}(2002){Press}, {Teukolsky}, {Vetterling}, \&
  {Flannery}}]{2002Press}
{Press}, W.~H., {Teukolsky}, S.~A., {Vetterling}, W.~T., \& {Flannery}, B.~P.
  2002, {Numerical recipes in C++ : the art of scientific computing}

\bibitem[{{Qiu} {et~al.}(2013){Qiu}, {Zhang}, {Menten}, {Liu}, \&
  {Tang}}]{2013Qiu}
{Qiu}, K., {Zhang}, Q., {Menten}, K.~M., {Liu}, H.~B., \& {Tang}, Y.-W. 2013,
  \apj, 779, 182

\bibitem[{{Qiu} {et~al.}(2014){Qiu}, {Zhang}, {Menten}, {Liu}, {Tang}, \&
  {Girart}}]{2014Qiu}
{Qiu}, K., {Zhang}, Q., {Menten}, K.~M., {et~al.} 2014, \apjl, 794, L18

\bibitem[{{Rao} {et~al.}(1998){Rao}, {Crutcher}, {Plambeck}, \&
  {Wright}}]{1998Rao}
{Rao}, R., {Crutcher}, R.~M., {Plambeck}, R.~L., \& {Wright}, M.~C.~H. 1998,
  \apjl, 502, L75

\bibitem[{{Rosolowsky} {et~al.}(2008){Rosolowsky}, {Pineda}, {Kauffmann}, \&
  {Goodman}}]{2008Rosolowsky}
{Rosolowsky}, E.~W., {Pineda}, J.~E., {Kauffmann}, J., \& {Goodman}, A.~A.
  2008, \apj, 679, 1338

\bibitem[{Roy {et~al.}(2011)Roy, Ade, Bock, Chapin, Devlin, Dicker, France,
  Gibb, Griffin, Gundersen, Halpern, Hargrave, Hughes, Klein, Marsden, Martin,
  Mauskopf, Morales~Ortiz, Netterfield, Noriega-Crespo, Olmi, Patanchon, Rex,
  Scott, Semisch, Truch, Tucker, Tucker, Viero, \& Wiebe}]{2011Roy}
Roy, A., Ade, P. A.~R., Bock, J.~J., {et~al.} 2011, \apj,
  727, 114

\bibitem[{{Rygl} {et~al.}(2012){Rygl}, {Brunthaler}, {Sanna}, {Menten}, {Reid},
  {van Langevelde}, {Honma}, {Torstensson}, \& {Fujisawa}}]{2012Rygl}
{Rygl}, K.~L.~J., {Brunthaler}, A., {Sanna}, A., {et~al.} 2012, \aap, 539, A79

\bibitem[Santos-Lima et al.(2010)]{2010Santos} Santos-Lima, R., Lazarian, A., de Gouveia Dal Pino, E.~M., \& Cho, J.\ 2010, \apj, 714, 442

\bibitem[{{Schneider} {et~al.}(2010){Schneider}, {Csengeri}, {Bontemps},
  {Motte}, {Simon}, {Hennebelle}, {Federrath}, \& {Klessen}}]{2010Schneider}
{Schneider}, N., {Csengeri}, T., {Bontemps}, S., {et~al.} 2010, \aap, 520, A49

\bibitem[{{Schneider} {et~al.}(2012){Schneider}, {Csengeri}, {Hennemann},
  {Motte}, {Didelon}, {Federrath}, {Bontemps}, {Di Francesco}, {Arzoumanian},
  {Minier}, {Andr{\'e}}, {Hill}, {Zavagno}, {Nguyen-Luong}, {Attard},
  {Bernard}, {Elia}, {Fallscheer}, {Griffin}, {Kirk}, {Klessen}, {K{\"o}nyves},
  {Martin}, {Men'shchikov}, {Palmeirim}, {Peretto}, {Pestalozzi}, {Russeil},
  {Sadavoy}, {Sousbie}, {Testi}, {Tremblin}, {Ward-Thompson}, \&
  {White}}]{2012Schneider}
{Schneider}, N., {Csengeri}, T., {Hennemann}, M., {et~al.} 2012, \aap, 540, L11

\bibitem[{{Schneider} {et~al.}(2016){Schneider}, {Bontemps}, {Motte},
  {Blazere}, {Andr{\'e}}, {Anderson}, {Arzoumanian}, {Comer{\'o}n}, {Didelon},
  {Di Francesco}, {Duarte-Cabral}, {Guarcello}, {Hennemann}, {Hill},
  {K{\"o}nyves}, {Marston}, {Minier}, {Rygl}, {R{\"o}llig}, {Roy}, {Spinoglio},
  {Tremblin}, {White}, \& {Wright}}]{2016Schneider}
{Schneider}, N., {Bontemps}, S., {Motte}, F., {et~al.} 2016, \aap, 591, A40

\bibitem[Shu et al.(1987)]{1987Shu} Shu, F.~H., Adams, F.~C., \& Lizano, S.\ 1987, \araa, 25, 23 

\bibitem[{{Shu}(1991)}]{1991Shu}
{Shu}, F. 1991, {Physics of Astrophysics, Vol. II: Gas Dynamics} (University
  Science Books)

\bibitem[{{Sridharan} {et~al.}(2014){Sridharan}, {Rao}, {Qiu}, {Cortes}, {Li},
  {Pillai}, {Patel}, \& {Zhang}}]{2014Sridharan}
{Sridharan}, T.~K., {Rao}, R., {Qiu}, K., {et~al.} 2014, \apjl, 783, L31

\bibitem[Stone et al.(1998)]{1998Stone} Stone, J.~M., Ostriker, E.~C., \& Gammie, C.~F.\ 1998, \apjl, 508, L99

\bibitem[{{Tang} {et~al.}(2009{\natexlab{a}}){Tang}, {Ho}, {Girart}, {Rao},
  {Koch}, \& {Lai}}]{2009aTang}
{Tang}, Y.-W., {Ho}, P.~T.~P., {Girart}, J.~M., {et~al.} 2009{\natexlab{a}},
  \apj, 695, 1399

\bibitem[{{Tang} {et~al.}(2009{\natexlab{b}}){Tang}, {Ho}, {Koch}, {Girart},
  {Lai}, \& {Rao}}]{2009bTang}
{Tang}, Y.-W., {Ho}, P.~T.~P., {Koch}, P.~M., {et~al.} 2009{\natexlab{b}},
  \apj, 700, 251

\bibitem[{{Tang} {et~al.}(2013){Tang}, {Ho}, {Koch}, {Guilloteau}, \&
  {Dutrey}}]{2013Tang}
{Tang}, Y.-W., {Ho}, P.~T.~P., {Koch}, P.~M., {Guilloteau}, S., \& {Dutrey}, A.
  2013, \apj, 763, 135

\bibitem[{{Tang} {et~al.}(2010){Tang}, {Ho}, {Koch}, \& {Rao}}]{2010Tang}
{Tang}, Y.-W., {Ho}, P.~T.~P., {Koch}, P.~M., \& {Rao}, R. 2010, \apj, 717,
  1262

\bibitem[{{Tassis} \& {Mouschovias}(2007)}]{2007TM}
{Tassis}, K., \& {Mouschovias}, T.~C. 2007, \apj, 660, 402

\bibitem[{Vall{\'e}e \& Fiege(2006)}]{2006VF}
Vall{\'e}e, J.~P., \& Fiege, J.~D. 2006, \apj, 636, 332

\bibitem[{{Van Loo} {et~al.}(2014){Van Loo}, {Keto}, \& {Zhang}}]{2014VanLoo}
{Van Loo}, S., {Keto}, E., \& {Zhang}, Q. 2014, \apj, 789, 37

\bibitem[{{Williams} {et~al.}(1994){Williams}, {de Geus}, \&
  {Blitz}}]{1994Williams}
{Williams}, J.~P., {de Geus}, E.~J., \& {Blitz}, L. 1994, \apj, 428, 693

\bibitem[{Wilner \& Welch(1994)}]{1994WW}
Wilner, D.~J., \& Welch, W.~J. 1994, ApJ, 427, 898

\bibitem[{Woody {et~al.}(1989)Woody, Scott, Scoville, Mundy, Sargent, Padin,
  Tinney, \& Wilson}]{1989Woody}
Woody, D.~P., Scott, S.~L., Scoville, N.~Z., {et~al.} 1989, ApJ, 337, L41

\bibitem[{{Zapata} {et~al.}(2009){Zapata}, {Ho}, {Schilke}, {Rodr{\'{\i}}guez},
  {Menten}, {Palau}, \& {Garrod}}]{2009Zapata}
{Zapata}, L.~A., {Ho}, P.~T.~P., {Schilke}, P., {et~al.} 2009, \apj, 698, 1422

\bibitem[{{Zapata} {et~al.}(2012){Zapata}, {Loinard}, {Su}, {Rodr{\'{\i}}guez},
  {Menten}, {Patel}, \& {Galv{\'a}n-Madrid}}]{2012Zapata}
{Zapata}, L.~A., {Loinard}, L., {Su}, Y.-N., {et~al.} 2012, \apj, 744, 86

\bibitem[{{Zhang} {et~al.}(1998){Zhang}, {Ho}, \& {Ohashi}}]{1998Zhang}
{Zhang}, Q., {Ho}, P.~T.~P., \& {Ohashi}, N. 1998, \apj, 494, 636

\bibitem[{{Zhang} {et~al.}(2015){Zhang}, {Wang}, {Lu}, \&
  {Jim{\'e}nez-Serra}}]{2015Zhang}
{Zhang}, Q., {Wang}, K., {Lu}, X., \& {Jim{\'e}nez-Serra}, I. 2015, \apj, 804,
  141

\bibitem[{Zhang {et~al.}(2009)Zhang, Wang, Pillai, \& Rathborne}]{2009Zhang}
Zhang, Q., Wang, Y., Pillai, T., \& Rathborne, J. 2009, ApJ, 696, 268

\bibitem[{{Zhang} {et~al.}(2014){Zhang}, {Qiu}, {Girart}, {(Baobab Liu},
  {Tang}, {Koch}, {Li}, {Keto}, {Ho}, {Rao}, {Lai}, {Ching}, {Frau}, {Chen},
  {Li}, {Padovani}, {Bontemps}, {Csengeri}, \& {Ju{\'a}rez}}]{2014Zhang}
{Zhang}, Q., {Qiu}, K., {Girart}, J.~M., {et~al.} 2014, \apj, 792, 116

\end{thebibliography}

\clearpage	

\begin{turnpage}
\begin{deluxetable}{ccccccccc}
\tabletypesize{\scriptsize}
\tablecaption{Observational Parameters\label{obs_table}}
\tablewidth{0pt}
\tablehead{
\multirow{2}{*}{Date} & \multirow{2}{*}{Configuration} & Number of & Polarimeter & \multicolumn{4}{c}{Calibrators} & \multirow{2}{*}{Source(s)} \\ \cline{5-8} & & Antennas & Mode & Gain & Flux & Bandpass & Leakage & }
\startdata
2011 Jul 09 & Subcompact & 7 & Single & MWC349A & Uranus & 3C279 & 3C279 & Cyg-N38 \\
2011 Jul 12 & Subcompact & 6 & Single & MWC349A & Uranus & 3C279 & 3C279 on 2011 Jul 13 & Cyg-N48, Cyg-N53 \\
2011 Jul 13 & Subcompact & 7 & Single & MWC349A & MWC349A & 3C279 & 3C279 & Cyg-N38, Cyg-N48, Cyg-N53 \\
2012 Jul 03 & Compact & 7 & Single & MWC349A & Titan & 3C279, 3C84 & 3C279 on 2012 Jul 04 & Cyg-N38, Cyg-N48 \\
2012 Jul 04 & Compact & 7 & Single & MWC349A & Titan & 3C279 & 3C279 & Cyg-N38, Cyg-N48 \\
2012 Jul 05 & Compact & 7 & Single & MWC349A & Titan & 3C279, 3C84 & 3C84 & Cyg-N38, Cyg-N48 \\
2012 Aug 07 & Subcompact & 6 & Single & MWC349A & Uranus & 3C279 & 3C84 on 2012 Aug 08 & Cyg-N38, Cyg-N48 \\
2012 Aug 08 & Subcompact & 6 & Single & MWC349A & Uranus & 3C84 & 3C84 & Cyg-N38, Cyg-N48 \\
2012 Aug 09 & Subcompact & 6 & Single & MWC349A & Uranus & 3C84 & 3C84 on 2012 Aug 08 & Cyg-N38, Cyg-N48 \\
2014 Jun 19 & Compact & 7 & Dual & MWC349A & Neptune & 3C279, 3C454.3 & 3C454.3 & Cyg-N43, Cyg-N51 \\
2014 Jun 21 & Compact & 7 & Dual & MWC349A & Neptune & 3C454.3 & 3C454.3 on 2014 Jun 19 & Cyg-N43, Cyg-N51 \\
2015 Jun 08 & Subcompact & 7 & Dual & MWC349A & Titan & 3C279 & 3C279 & Cyg-N53 \\
2015 Jun 12 & Subcompact & 7 & Dual & MWC349A & Neptune & 3C279, 3C454.3 & 3C279 & Cyg-N43 \\
2015 Jun 19 & Subcompact & 7 & Dual & MWC349A & Titan & 3C279 & 3C279 & Cyg-N51 \\
\enddata
\end{deluxetable}
\end{turnpage}
\clearpage

\begin{deluxetable}{lcccccccccc}
\tabletypesize{\scriptsize}
\tablecaption{Mapping Parameters\label{map_table}}
\tablewidth{0pt}
\tablehead{ 
\multirow{2}{*}{Source} & \multicolumn{2}{c}{Pointing Center} & & On-source Time & \multirow{2}{*}{} & \multicolumn{2}{c}{Synthesized Beam} & \multirow{2}{*}{} & rms Noise & \multirow{2}{*}{Other Name} \\
\cline{2-3} \cline{5-5} \cline{7-8} \cline{10-10} & $\alpha$ (J2000) & $\delta$ (J2000) & & Single/Dual (hour) & & HPBW(\arcsec) & P.A. (\arcdeg) & & Stokes $I$/Pol (mJy beam$^{-1}$)}
\startdata
Cyg-N38 & 20:38:59.11 & 42:22:25.96 & & 9.4/0 & & 3.6 $\times$ 2.9 & 65 & & 15/1.6 & DR21(OH)-W \\
Cyg-N43 & 20:39:00.60 & 42:24:34.99 & & 0/3.1 & & 4.6 $\times$ 3.5 & 55 & & 10/1.6 & W75S-FIR1 \\
Cyg-N44$^a$ & 20:39:01.20 & 42:22:48.50 & & 4.7/0 & & 3.9 $\times$ 3.4 & 56 & & 20/3.1 & DR21(OH) \\
Cyg-N48 & 20:39:01.34 & 42:22:04.89 & & 7.5/0 & & 3.3 $\times$ 3.0 & 63 & & 13/1.7 & DR21(OH)-S \\
Cyg-N51 & 20:39:02.40 & 42:24:59.00 & & 0/3.5 & & 4.4 $\times$ 3.4 & 54 & & 8.9/1.6 & W75S-FIR2 \\
Cyg-N53 & 20:39:02.96 & 42:25:50.99 & & 1.5/2.1 & & 4.4 $\times$ 3.0 & 60 & & 14/1.9 & -- \\
\enddata
\tablenotetext{a}{The map of SMA subcompact, compact and extended data in \citet{2013Girart}.}
\end{deluxetable}

\begin{turnpage}
\begin{deluxetable}{ccccccccccccc}
\tabletypesize{\scriptsize}
\tablecaption{Physical Parameters of the Cores}
\tablewidth{0pt}
\tablehead{ 
\multirow{2}{*}{Source} & S$_{int}$ & \multicolumn{3}{c}{FWHM$^a$} &&  \multicolumn{4}{c}{Postion Angle} & Mass & N$_{H_2}$ &  n$_{H_2}$ \\
\cline{3-5} \cline{7-10}
 & (Jy) & Major ($\arcsec$) & Minor ($\arcsec$) & Mean$^b$ (mpc) && $\theta_{core}$ ($\arcdeg$) & $\theta_{JCMT}^c$ ($\arcdeg$) & $\theta_{SMA}^c$ ($\arcdeg$) & $\theta_{vg}$ ($\arcdeg$) & (M$_{\sun}$) & (10$^{23}$ cm$^{-2}$) & (10$^{6}$ cm$^{-3}$)}
\startdata
Cyg-N38 & 4.3 $\pm$ 0.3 & 11.5 $\pm$ 0.4 & 7.8 $\pm$ 0.6 & 64 $\pm$ 3 && -27.0 $\pm$ 3.7 & 85.0 $\pm$ 5.1 & -50.0 $\pm$ 36.2 & -- & 59 $\pm$ 4 & 2.2 $\pm$ 0.2 & 1.1 $\pm$ 0.1 \\
Cyg-N43 & 1.2 $\pm$ 0.1 & 14.3 $\pm$ 0.3 & 5.7 $\pm$ 0.4 & 61 $\pm$ 2 && -74.4 $\pm$ 1.7 & -44.5 $\pm$ 24.4 & -24.0 $\pm$ 38.1 & $\sim$ -31 & 16 $\pm$ 1 & 0.7 $\pm$ 0.1 & 0.4 $\pm$ 0.03 \\
Cyg-N44 & 21.4 $\pm$ 0.3 & 11.1 $\pm$ 0.5 & 9.4 $\pm$ 0.1 & 69 $\pm$ 2 && 13.3 $\pm$ 3.7 & 78.2 $\pm$ 5.9 & 47.6 $\pm$ 30.1 & $\sim$ 63 & 167 $\pm$ 2 & 5.3 $\pm$ 0.2 & 2.5 $\pm$ 0.1 \\
Cyg-N44 MM1 & 2.0 $\pm$ 0.1 & $<$ 4.3 $\pm$ 0.1 & $<$ 3.2 $\pm$ 0.1 & $<$ 25 $\pm$ 1 && 52.3 $\pm$ 1.3 & -- & -- & -- & 16 $\pm$ 1 & $>$ 3.8 $\pm$ 0.2 & $>$ 4.8 $\pm$ 0.4 \\
Cyg-N44 MM2 & 0.2 $\pm$ 0.1 & $<$ 2.5 $\pm$ 0.1 & $<$ 2.1 $\pm$ 0.1 & $<$ 16 $\pm$ 1 && 16.0 $\pm$ 9.6 & -- & -- & -- & 2 $\pm$ 1 & $>$ 1.0 $\pm$ 0.5 & $>$ 2.1 $\pm$ 0.1 \\
Cyg-N48 & 7.5 $\pm$ 0.3 & 14.3 $\pm$ 0.2 & 7.2 $\pm$ 0.4 & 69 $\pm$ 2 && -71.5 $\pm$ 1.0 & -59.2 $\pm$  9.3 & -60.8 $\pm$ 44.0 & $\sim$ -72 & 102 $\pm$ 4 & 3.3 $\pm$ 0.2 & 1.5 $\pm$ 0.1 \\
Cyg-N51 & 3.3 $\pm$ 0.2 & 16.7 $\pm$ 0.3 & 7.4 $\pm$ 0.3 & 75 $\pm$ 2 && 7.5 $\pm$ 0.7 & -84.9 $\pm$ 8.9 & 89.3 $\pm$ 35.3 & -- & 45 $\pm$ 3 & 1.2 $\pm$ 0.1 & 0.5 $\pm$ 0.03 \\
Cyg-N53 & 1.9 $\pm$ 0.1 & 4.0 $\pm$ 0.3 & 2.6 $\pm$ 0.1 & 22 $\pm$ 1 && -14.3 $\pm$ 3.4 & 73.7 $\pm$ 22.2 & -23.5 $\pm$ 18.3 & -- & 26 $\pm$ 1 & 8.3 $\pm$ 0.2 & 12.2 $\pm$ 0.2 \\
\enddata
\label{phy_table}
\tablenotetext{a}{Deconvolved FWHM; for the unresolved sources Cyg-N44 MM1 and MM2, not deconvolved FWHMs are listed}
\tablenotetext{b}{Geometrical mean; $FWHM_{mean}$ = $\sqrt{FWHM_{major} \times FWHM_{minor}}$}
\tablenotetext{c}{Magnetic field direction}
\end{deluxetable}
\end{turnpage}
\clearpage

\begin{deluxetable}{cccccccc}
\tabletypesize{\scriptsize}
\tablecaption{Angular Dispersion Function Fit Parameters}
\tablewidth{0pt}
\tablehead{ 
\multirow{2}{*}{Source} & $\delta$ & \multirow{2}{*}{$\langle B_t^2 \rangle / \langle B_0^2 \rangle $} & $a_2^\prime$ & \multirow{2}{*}{N} & $\delta V_{lsr}$ & $B_0$ & $B_{total}$ \\
 & (mpc) & & (10$^{-3}$ arcsec$^{-2}$) & & (km s$^{-1}$) & (mG) & (mG)}
\startdata
Cyg-N38 & 26 $\pm$ 12 & 3.8 $\pm$ 2.4 & -1.9 $\pm$ 1.6 & 4.7 $\pm$ 0.1 & 1.5 $\pm$ 0.2 & 0.56 $\pm$ 0.19 & $\sim$ 0.69 \\
Cyg-N43 & 26 $\pm$ 31 & 2.8 $\pm$ 5.3 & 2.5 $\pm$ 6.6 & 4.5 $\pm$ 0.3 & 1.6 $\pm$ 0.4 & 0.42 $\pm$ 0.41 & $\sim$ 0.51 \\
Cyg-N44 & 22 $\pm$ 5 & 2.0 $\pm$ 0.2 & 0.8 $\pm$ 0.4 & 5.2 $\pm$ 0.5 & 2.2 $\pm$ 0.7 & 1.71 $\pm$ 0.55 & $\sim$ 1.75 \\
Cyg-N48 & 15 $\pm$ 3 & 4.5 $\pm$ 1.2 & 0.1 $\pm$ 0.4 & 7.4 $\pm$ 2.2 & 1.2 $\pm$ 0.2 & 0.48 $\pm$ 0.10 & $\sim$ 0.59 \\
Cyg-N51 & 22 $\pm$ 6 & 2.9 $\pm$ 0.6 & 0.1 $\pm$ 0.7 & 5.7 $\pm$ 0.6 & 1.6 $\pm$ 0.2 & 0.46 $\pm$ 0.08 & $\sim$ 0.56 \\
DR21 Filament$^a$ & 151 $\pm$ 21 & $\sim$ 0.16 & -- & $\sim$ 1 & $\sim$ 0.8 & $\sim$ 0.62 & $\sim$ 0.94 \\
\enddata
\label{phy_table2}
\tablenotetext{a}{Values in \citet{2013Girart}} 
\end{deluxetable}

\begin{deluxetable}{ccccccccccc}
\tabletypesize{\scriptsize}
\tablecaption{Comparison of the Kinematic, Magnetic, and Gravitational Energies}
\tablewidth{0pt}
\tablehead{ 
\multirow{2}{*}{Source} & $c_s$ & $\sigma_{v}$ & $V_A$ & \multirow{2}{*}{$m_s$} & \multirow{2}{*}{$m_A$} & \multirow{2}{*}{$\beta$} & $M/\Phi_B$ & \multirow{2}{*}{$|\frac{\mathcal{T}}{\mathcal{W}}|$} & \multirow{2}{*}{$|\frac{\mathcal{M}}{\mathcal{W}}|$} & \multirow{2}{*}{$|\frac{2\mathcal{T}+\mathcal{M}}{\mathcal{W}}|$}\\
 &  (km s$^{-1}$) &  (km s$^{-1}$)&  (km s$^{-1}$) & & & & ($1/2\pi\sqrt{G}$) & & & }
\startdata
Cyg-N38 & 0.34 & 1.5 & 0.9 & 7.5 & 2.7 & 0.27 & 2.4 & 1.4 & 0.19 & 3.0 \\
Cyg-N43 & 0.34 & 1.6 & 1.2 & 8.0 & 2.4 & 0.18 & 1.0 & 5.7 & 0.99 & 12.4\\
Cyg-N44 & 0.42 & 2.2 & 1.6 & 9.0 & 2.4 & 0.14 & 2.3 & 1.2 & 0.20 & 2.6\\
Cyg-N48 & 0.34 & 1.2 & 0.7 & 6.0 & 3.0 & 0.50 & 4.3 & 0.6 & 0.06 & 1.3\\
Cyg-N51 & 0.34 & 1.6 & 1.1 & 8.0 & 2.4 & 0.18 & 1.6 & 2.5 &0.42 & 5.4\\
DR21 Filament & 0.30$^a$ & 0.8 & 3.0$^b$ & 4.6 & 0.5 & 0.02 & 3.4$^c$ & 0.04 & 0.28 & 0.36\\
\enddata
\label{phy_table3}
\tablenotetext{a}{Assume 15 K from the dust temperature map of \citet{2012Hennemann}}
\tablenotetext{b}{Adopt $n_{H_2} = 2 \times 10^{5} cm^{-3}$ \citep{2013Girart}}
\tablenotetext{c}{Adopt $N_{H_2} = 41.6 \times 10^{22} cm^{-2}$ \citet{2012Hennemann}}
\end{deluxetable}
\clearpage

\clearpage

\clearpage

\clearpage

\clearpage

\clearpage

\clearpage

\clearpage

\clearpage

\clearpage

\clearpage

\global\pdfpageattr\expandafter{\the\pdfpageattr/Rotate 90}
\end{document}